\documentclass[twocolumn,reprint,nofootinbib,longbibliography,prd,superscriptaddress]{revtex4-2}
\usepackage{float}
\usepackage{tikz,pgfplots,soul}
\usepackage{subfigure}
\usepackage{amsfonts, amsmath, amssymb, bm,  enumerate, float, graphicx, graphics, color,mathrsfs,hyperref,nicefrac,physics,revsymb}
\usepackage[utf8]{inputenc}

\newcommand{\td}{\text{d}}

\newcommand{\be}{\begin{equation}}
\newcommand{\ee}{\end{equation}}
\newcommand{\bea}{\begin{eqnarray}}
\newcommand{\eea}{\end{eqnarray}}

\newcommand{\uh}{\underline{h}}

\newcommand{\sD}{\mathcal{D} \mspace{-11mu} \slash}

\setuldepth{C}

\definecolor{ktbgreen}{RGB}{0, 100, 0}

\begin{document}
		
		%\title{Horizon Evolution when Departing from a Moment of Time Symmetry}
\title{Marginally Outer Trapped Tori in Black Hole Spacetimes}

\author{Kam To Billy Sievers}
\email{sieversktb@mcmaster.ca}
\affiliation{Department of Mathematics and Statistics and Department of Physics and Astronomy, McMaster University, Hamilton, Ontario, L8S 4M1, Canada}

\author{Liam Newhook}
\email{liamn@mun.ca}
\affiliation{Department of Physics and Physical Oceanography, Memorial University,
St. John’s, Newfoundland and Labrador, A1B 3X7, Canada}

\author{Sarah Muth}
\email{smmmuth@mun.ca}
\affiliation{Department of Mathematics and Statistics, Memorial University of Newfoundland, St. John's, Newfoundland and Labrador, A1C 5S7, Canada}

\author{Ivan Booth}
\email{ibooth@mun.ca}
\affiliation{Department of Mathematics and Statistics, Memorial University of Newfoundland, St. John's, Newfoundland and Labrador, A1C 5S7, Canada}

\author{Robie A. Hennigar}
\email{robie.hennigar@icc.ub.edu}
\affiliation{Departament de F{\'\i}sica Qu\`antica i Astrof\'{\i}sica, Institut de
Ci\`encies del Cosmos,
 Universitat de
Barcelona, Mart\'{\i} i Franqu\`es 1, E-08028 Barcelona, Spain
}

\author{Hari K. Kunduri}
\email{kundurih@mcmaster.ca}
\affiliation{Department of Mathematics and Statistics and Department of Physics and Astronomy, McMaster University, Hamilton, Ontario, L8S 4M1, Canada}

		\begin{abstract}
            During a binary black hole merger, 
 multiple intermediary marginally outer trapped tubes connect the initial pair of apparent horizons with the final (single) 
 apparent horizon. The marginally outer trapped surfaces (MOTSs) that foliate these tubes can have complicated geometries as well as non-spherical topologies. In particular, toroidal MOTSs form inside both of the original black holes during the early stages of a head-on merger that starts from time-symmetric initial data~\cite{Pook-Kolb:2021jpd}.  We show that toroidal MOTSs also form in the maximal analytic extension of the Schwarzschild spacetime as Kruskal time advances from the $T=0$ moment of time symmetry. As for the merger simulations, they cross the Einstein-Rosen bridge and are tightly sandwiched between the apparent horizons in the two asymptotic regions at early times.  This strongly suggests that their formation is a consequence of the initial conditions rather 
   than merger physics. 
   Finally, we consider MOTSs of spherical topology in the Kruskal-Szekeres slicing and study their properties. All of these are contained within the apparent horizon but some do not enclose the wormhole.
		\end{abstract}
		
		\maketitle

\section{Introduction}

Horizon evolution during a binary black hole merger has been studied for more than five decades. For event horizons, the qualitative picture of how two black 
holes become one has been understood
since at least the early 1970s \cite{hawking_ellis_1973,Hawking:1972hy},\footnote{See also~\cite{Gadioux:2023pmw} for recent work on this subject.} but while some knowledge of apparent horizon 
mergers dates to the same time \cite{Hawking:1972hy,CADEZ1974449}, it is only recently that
the more intricate process for apparent 
horizons has been carefully studied in numerical simulations
\cite{Moesta:2015sga,PhysRevLett.123.171102,pook-kolb2020dynamical,PhysRevD.100.084044,pookkolb2020horizons,Pook-Kolb:2021gsh,Pook-Kolb:2021jpd,Booth:2021sow}.
Event horizon mergers are 
described by the fairly simple ``pair-of-pants'' diagram \cite{hawking_ellis_1973},
however the process by which two distinct apparent horizons
become one is significantly more complicated and involves multiple
marginally outer trapped surfaces (MOTSs) with complicated (often self-intersecting) geometries. Identifying these surfaces has required
the introduction of new MOTS-finding techniques \cite{pook-kolb:2018igu,Booth:2020qhb,Booth:2021sow}.

% The evolution of 
% these 
% surfaces includes pair creations and annihilations 
% with an intricate (possibly infinite) series of these 
% events midwifing the birth of final 

% that the two original apparent horizons ultimately 
% disappear leaving behind the apparent horizon of the final, merged, black hole. 

While studying the full horizon evolution during a merger necessarily 
involves numerical simulations, it turns out that some 
aspects of the process can also be studied with exact solutions.
Notably, self-intersecting MOTSs, which were first observed in numerical simulations \cite{PhysRevLett.123.171102}, have subsequently been 
found to be very common, including in many 
exact black hole solutions \cite{Booth:2020qhb,Hennigar:2021ogw,Booth:2022vwo}. Following from 
the example of event horizons \cite{Emparan:2016ylg,Emparan:2017vyp}, there has also 
been an attempt to understand the horizon dynamics of extreme 
mass ratio mergers using the pure Schwarzschild spacetime 
\cite{Booth:2020qhb}. 

\begin{figure}    \includegraphics{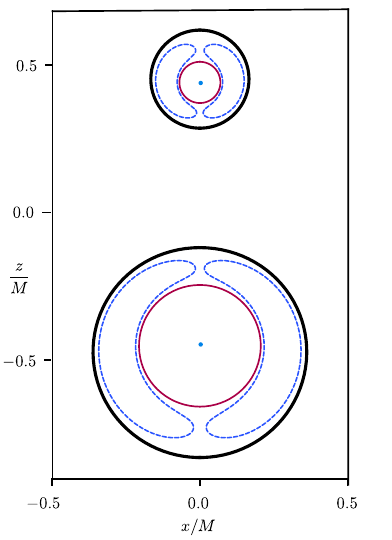}
    \caption{During the early stages of a head-on black hole merger toroidal MOTSs (the blue dashed lines) were observed to form between the (black) outermost MOTSs (the apparent horizons) and the (red) inner MITSs (apparent horizons relative to the asymptotic regions on the other sides of the wormhole).  The central (light blue) points are $\infty$ in the ``internal'' asymptotic regions. This is adapted from Figure 9, $T=M$ in \cite{Pook-Kolb:2021jpd}. See that paper for details of the simulation.% but note that the coordinate representation used in the figure is a fairly good representation of the geometry: hence at this point in time the apparent horizons are still quite close to being geometric spheres. 
    }
    \label{fig:numtor}
\end{figure}

\begin{figure}
    \centering
    \includegraphics[scale=0.4]{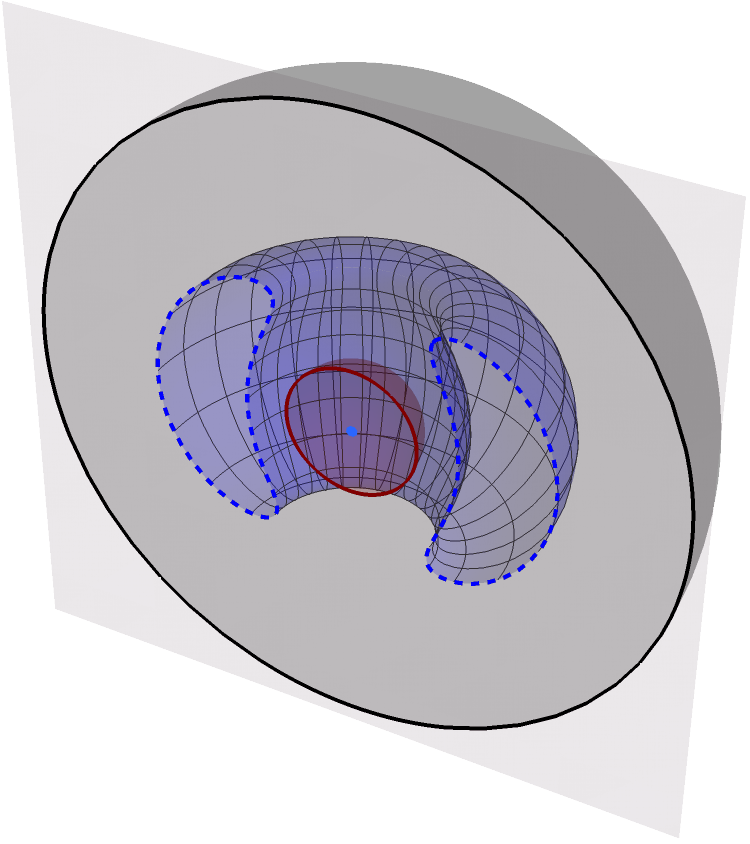}
    \caption{Three-dimensional cross-section showing the 
    equatorial cross-sections from Figure \ref{fig:numtor} 
    rotated into tori. As usual the light-blue dot in the centre is actually $\infty$ in the internal asymptotic region. }
    \label{fig:DonutRotated}
\end{figure}

In this paper, we return to the Schwarzschild spacetime to investigate another phenomenon that was first observed
in numerical simulations.  During the merger simulations of \cite{Pook-Kolb:2021jpd}, toroidal MOTSs were observed inside both 
of the original apparent horizons. See Figure \ref{fig:numtor} for an equatorial cross-section of a snapshot of this simulation 
and then Figure \ref{fig:DonutRotated} for how the blue kidney-shaped MOTSs rotate into tori. 
However, as is evident from the figure, these tori existed at a time when the two black holes were still fairly well separated and not exhibiting strong gravitational distortions.\footnote{Toroidal MOTSs were also present at earlier times in the simulation. 
However at those earlier times, they are harder to visually distinguish.}
As such it seemed likely that the toroidal MOTSs were not a consequence of merger physics. 

% \begin{figure}
%     \includegraphics[width=.5
% \textwidth]{figs/donut3.pdf}
%     \caption{
%     A 3-dimensional visualization of the toroidal MOTS. The right is a $y=0$ cross-section (the shaded plane) of the objects on the left found at $T=0.8$. The blue curve is the cross-section of the toroidal MOTS and the black curves are those of the spherical horizons found at $X^2=T^2$.}
%     \label{fig:toroidal_3d}
% \end{figure}

\begin{figure}
    \centering
    \includegraphics{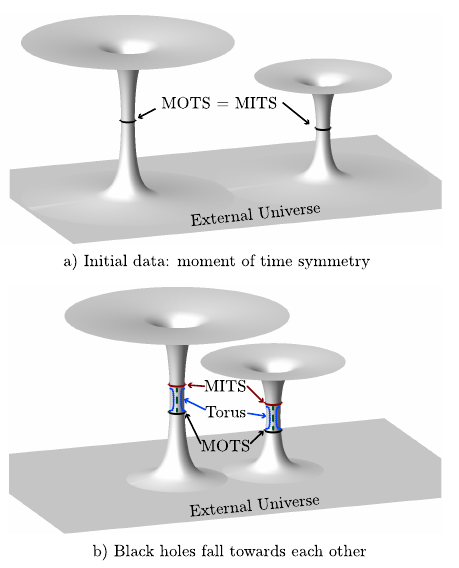}
    \caption{Cartoon of the early stages of a black hole merger, departing from a moment of time-symmetry. a) represents the initial conditions: a moment of time symmetry in which the horizons from both asymptotic regions coincide. b) is after some time evolution when the horizons have separated and the toroidal MOTS have appeared. Dashed green lines are the
``north poles'' of the black holes.}
    \label{fig:EmbeddingPicture}
\end{figure}

Instead, we suspected that they were a by-product of the initial conditions. These simulations were started from  Brill-Lindquist initial data. This is time symmetric and avoids the black hole singularities by using time slices that extend through the Einstein-Rosen throats of both black holes into the ``universes'' on the other side of the wormholes \cite{PhysRev.131.471}. This is depicted in Figure \ref{fig:EmbeddingPicture}a) with the usual asymptotic region on the bottom and the two internal ones on top. As a consequence of the time symmetry, MOTSs are minimal surfaces of the original time-slice intrinsic geometry and so are also marginally inner trapped surfaces (MITSs). Equivalently, each is marginally outer trapped with respect to both the top and bottom asymptotic regions (which in this case serve to distinguish between outward and inward directions). This is also shown in the figure. 

Once the evolution begins and the spacetime evolves away from the initial moment of time symmetry, the degeneracy is lost and the two minimal surfaces split into a distinct MOTS and MITS (or equivalently two MOTSs with one facing the external and one the internal asymptotic regions). These are on opposite sides of the narrowest part of the  wormhole throat and the toroidal MOTS lies in between them, straddling the throat. All of this is shown in Figure \ref{fig:EmbeddingPicture}b). Note that unlike embedding diagrams commonly seen in introductory textbooks, this is not a surface of constant radial coordinate in the equatorial
plane ($\theta=\pi/2$). Instead it is a double-copy of the $\phi=0$ and $\phi=\pi$ meridians with $0 < \theta < \pi$. 
%\sout{this is a non-standard embedding diagram since one is not restricted to an equatorial plane $\theta = \pi/2$ with a radial and azmuthal angle displayed. Rather, here was embed with $r$ and a double-copy of the 
%polar angle $\theta$ (i.e., we extend the range of $\theta$ from $(0,\pi)$ to $(0,2\pi)$) For definiteness one can think of this as embedding both $\phi = 0, \pi$, and connecting them at the north and south pole.} 
To emphasize this point, the north pole $\theta = 0$ is shown as a dotted green line in the figure (and the south pole is invisible on the other side).  

The Schwarzschild spacetime written in Kruskal-Szekeres coordinates can be used to test the idea that 
toroidal MOTSs are a by-product of the departure from time-symmetry, rather than a core property of black hole mergers. 
The spacelike hypersurface at timelike coordinate $T=0$ in this coordinate system is a moment of time symmetry (and is an example of a Brill-Lindquist
spacetime with one black hole). However, the surfaces for other values of $T$ are no 
longer time symmetric and so we can use this simplest black hole spacetime
to explore the departure from time symmetry. 

This paper is then organized in the following way. In Section~\ref{sec:Kruskal} we collect a number of results that will be crucial for the analysis in the manuscript. We review the Kruskal-Szekeres slicing of the Schwarzschild space-time, derive the equations to determine MOTSs in this slicing, discuss the methods used for visualizing MOTSs, the methods for determining the topology of a MOTS, and discuss the stability operator and the pseudo-spectral numerical techniques we use to obtain its eigenvalue spectrum. Then, we present our results in a streamlined fashion. First, in Section~\ref{sec:toroidal} we analyse the MOTSs of toroidal topology found in the Kruskal-Szekeres slicing. These MOTSs are the main result of our work. In Section~\ref{sec:spherical} we discuss additional MOTSs with spherical topology, including those with and without self-intersections. 

%\ivan{Note= sure where this fits in (maybe just mention it in the outline and then do it later?:
%%There is also a mathematical relativity reason to study these toroidal MOTS\footnote{We would %like to thank Graham Cox for pointing this out to us.}.}

Note that while we believe that \cite{pook_kolb_daniel_2021_4687700} represents the first observation of toroidal
MOTSs in black hole merger spacetimes, it was certainly not the first time that toroidal MOTSs or event horizons
have appeared in the literature. For example, \cite{Newman_1987} constructed spacetimes satisfying the 
dominant energy condition and containing MOTSs of a variety of topologies, \cite{Husa:1996xz} constructed 
time-symmetric vacuum initial data that contained toroidal MOTSs, \cite{Flores_2010,Mach:2017peu}
identified toroidal MOTS in closed FLRW spacetimes and \cite{Karkowski:2017sqk} constructed time symmetric
non-vacuum initial that contained toroidal MOTSs. Furthermore, \cite{Shapiro:1995rr, Bohn:2016soe} observed
event horizons with toroidal cross-sections during black hole mergers (though these are, of course, not MOTSs).

\section{General Considerations}
\label{sec:Kruskal}

To streamline the presentation, here we collect all the necessary preliminary details used in our analysis.

\subsection{Intrinsic and extrinsic geometry of a two-surface $\mathcal{S}$}
\label{Sec:Geom}

Let $(\mathcal{S}, q_{AB}, \mathcal{D}_A)$ be a spacelike two-surface embedded in a four-dimensional spacetime 
$(M, g_{ab}, \nabla_a)$. The metric on $\mathcal{S}$ is induced by the full spacetime metric:
\begin{equation}
    q_{AB} = e_A^a e_B^b g_{ab}
\end{equation}
where $e_A^a$ is the pull-back operator. 

The normal space at any point on $\mathcal{S}$ can be spanned by a pair of null normal vectors $\ell_+$ and $\ell_-$. We assume that these can be extended to smooth vector fields over 
$\mathcal{S}$ and for purposes of this paper scale them so that 
\begin{equation}
    \ell_+ \cdot \ell_- = -2 \; . \label{eq:scaling}
\end{equation}
Then the inverse metric can be written as
\begin{equation}
    g^{ab} = e^a_A e^b_B q^{AB} - \frac{1}{2} \left(\ell_+^a \ell_-^b - \ell_+^b \ell_-^a \right) 
\end{equation}

Derivatives of the null vectors over the surface characterize the extrinsic geometry of $\mathcal{S}$:
\begin{align}
    e_A^a e_B^b \nabla_a \ell^{\pm}_b & = \frac{1}{2} \theta_{\pm} q_{AB} + \sigma^{\pm}_{AB}
\end{align}
where $\theta_{\pm} := q^{ab} \nabla_a \ell^{\pm}_b$ are the traces of the left-hand quantities and the $\sigma^{\pm}_{AB}$ are the tracefree parts. We assume that $\mathcal{S}$ has an identified inside and outside and further that $\ell_-$ points in while $\ell_+$ points out. Then $\theta_{\pm}$ are respectively the expansion of congruences
of null curves that cross $\mathcal{S}$ tangent to $\ell_{\pm}$ while $\sigma^{\pm}_{AB}$ are the shears. 
Such surfaces are \emph{outer-trapped} if $\theta_+ < 0$ and \emph{marginally outer trapped} if $\theta_+=0$. 

Finally, the H\'aji\v{c}ek one-form\cite{Hajicek:1974oua}
\begin{equation}
    \omega_B = - \frac{1}{2} e_B^b \ell^-_a \nabla_b \ell_+^a 
\end{equation} 
is the connection on the normal bundle to $\mathcal{S}$. Under rescalings of the null vectors 
$\tilde{\ell}_{\pm} = e^{\pm \gamma} \ell_{\pm}$ it transforms as
\begin{equation}
    \tilde{\omega}_A = \omega_A + \mathcal{D}_A \gamma \; . 
\end{equation}

For  now we are chiefly interested in $\theta_+$, but the shear and the
H\'aji\v{c}ek one-form will return in Section \ref{sec:stab}
when we consider the stability 
operator.

\subsection{Kruskal-Szekeres Coordinates}

The well known maximal analytic extension of the Schwarzschild exterior solution is the Kruskal geometry
\begin{equation}
\td s^2 = N^2 (-\td T^2 + \td X^2) + r^2 \td \Omega^2,
\end{equation} 
where $T, X \in \mathbb{R}$ and $\td \Omega^2 = \td \theta^2 + \sin^2 \!\theta \,\td \phi^2$ is the standard unit round metric on $S^2$, with $\theta \in (0,\pi)$ and $\phi \sim \phi + 2\pi$. The lapse $N$ is defined by
\begin{equation}
   N^2 :=  \frac{32 M^3 e^{-\frac{r}{2M}}}{r} \; ,  \label{eqn:N}
\end{equation}
while $r>0$ is defined implicitly in terms of $T$ and $X$ by
\begin{equation}
T^2 - X^2 = \left(1 - \frac{r}{2M}\right) e^{\frac{r}{2M}}. \label{eqn:TXr} 
\end{equation}
The explict solution to this equation is 
\begin{equation}
r(T,X) = 2M \left[ 1 + W\left(\frac{-T^2 + X^2}{e}\right) \right] \label{eqn:r} \, ,  
\end{equation} 
where $W$ is the Lambert-$W$ function.

Note that, in contrast to the region outside the event horizon covered by Schwarzschild coordinates, 
the full Schwarzschild spacetime is not static. In particular, the timelike vector field $\partial_T$ is not Killing and it is clear that the metric components depend on $T$. %and, as noted, the induced metric on 
%the $\Sigma_T$ is time dependent. 
However, there still remains one moment of time symmetry: $T=0$.

\begin{figure}[h]
    \centering
    \includegraphics[width=.35\textwidth]{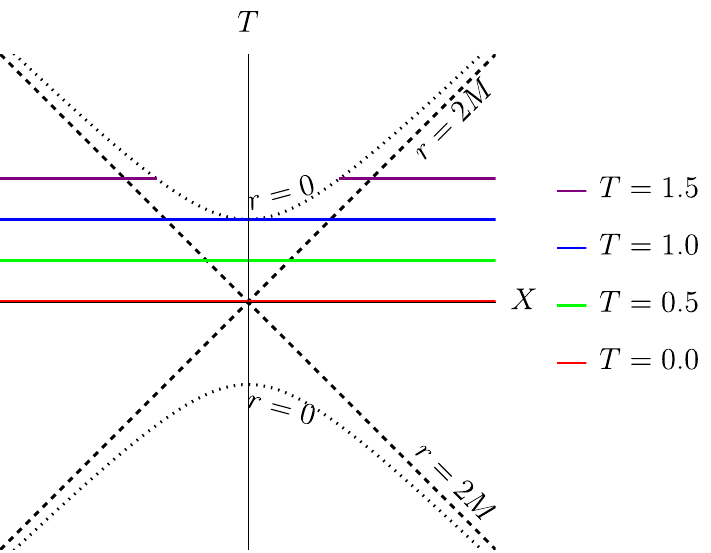}
\caption{A Kruskal-Szekeres diagram highlighting four constant $T$ slices, which correspond to Figures \ref{fig:penrosediagrams1} and \ref{fig:embedded1}. The dashed line is the event horizon $r=2M$, the dotted line is the curvature singularity at $r=0$. The solid lines are the $T,X$ axes.}
    \label{fig:kruskalDiagram}
\end{figure}

\begin{figure}[h]
    \centering
    \includegraphics[width=.49\textwidth]{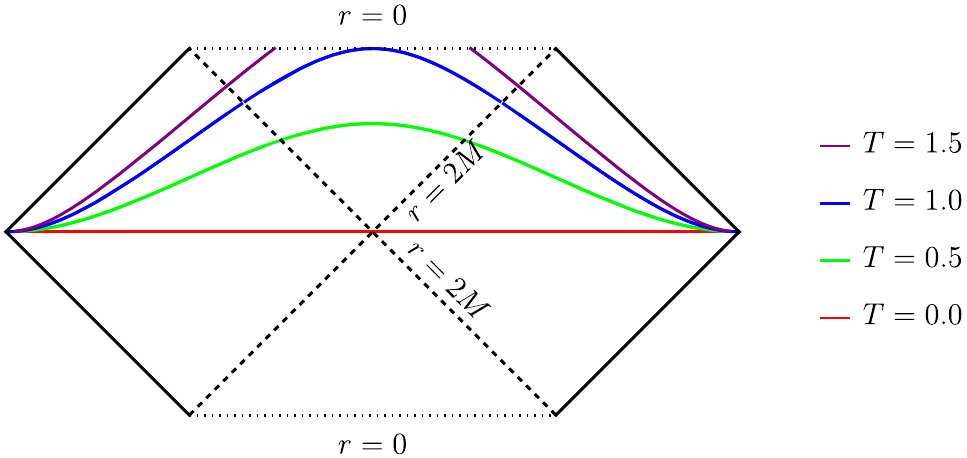}
\caption{A typical Carter-Penrose diagram of the maximally extended Schwarzschild spacetime. Lines correspond to the same lines in Figure \ref{fig:kruskalDiagram}. For $|T|\leq 1$, the $\Sigma_T$ have two asymptotically flat ends: $i_1^o$ and $i_2^o$. }
    \label{fig:penrosediagrams1}
\end{figure}

The Schwarzschild geometry can be represented in Kruskal-Szekeres diagrams such as in Figure \ref{fig:kruskalDiagram}, where the vertical and horizontal axes correspond to the $T,X\in\mathbb{R}$ coordinates, respectively. On these diagrams, constant $t,r$ slices are hyperbolic with the singularity at $r=0$ corresponding to $T^2-X^2=1$. The surfaces of constant Kruskal time can equally well be represented on the familiar Schwarzschild Carter-Penrose diagram, which we show in Figure \ref{fig:penrosediagrams1}. For $T^2 < 1$, the surfaces of constant time form Einstein-Rosen bridges (wormholes) connecting the left and right asymptotically flat regions. When $T^2=1$, the throat of the wormhole touches the singularity at the symmetric point $X = 0$. For $T^2 > 1$, the surfaces of constant time are comprised of two disconnected components terminating at $X = \pm \sqrt{T^2-1}$. These ideas are concretely visualized in the embedding diagrams such as Figure \ref{fig:embedded1}.

% The coordinates exhibit the $r=0$ singularity only for constant $T$ slices of $T^2\geq 1$. Otherwise, $T^2<1$ slices show a wormhole (Einstein-Rosen bridge) between $X<0$ and $X>0$ regions.\ktb{could talk about how these are intransversable objects, need citations}. As $T^2\to 1$, the wormhole throat `closes', before pinching off at the $T^2=1$ slice. $T^2>1$ slices then show two regions, or `universes', at $X>\sqrt{T^2-1}$ and $X<\sqrt{T^2-1}$. These ideas are concretely visualized in the embedding diagrams such as Figure \ref{fig:embedded1}.

\subsection{MOTS in the Kruskal-Szekeres time slices}

Given a foliation of a spacetime $M$ into spacelike three-surfaces $(\Sigma_T,h_{ij}, D_i)$ and a
two-surface $\mathcal{S}$ embedded in one $\Sigma_T$, a natural (though certainly not unique) scaling of the 
null normals to $\mathcal{S}$ is
\begin{equation}
    \ell_+ = \hat{u} + \hat{N} \quad \mbox{and} \quad \ell_- = \hat{u} - \hat{N} 
\end{equation}
where $\hat{u}$ is the forward-in-time pointing normal to the $\Sigma_T$  and $\hat{N}$ is the outward pointing unit spacelike normal to $\mathcal{S}$ in $\Sigma_T$\footnote{It is in order to make this choice that we are choosing 
the scaling $\ell_+ \cdot \ell_- = -2$.}. 

With this scaling the outward null expansion can be written as 
\begin{align}
    \theta_+ = q^{ij} K_{ij} + q^{ij} D_i \hat{N}_j \label{eq:MOTS}
\end{align}
where $K_{ij} = e_i^a e_j^b \nabla_a \hat{u}_b $ is the extrinsic curvature of $\Sigma_T$ in $M$ and 
$q^{ij} = e^i_A e^j_B q^{AB}$ is the push-forward of the inverse metric on $\mathcal{S}$ into $\Sigma_T$. Then our 
goal in this section is to solve for $\mathcal{S}$ with $\theta_+=0$. To do this, we 
apply the formalism of \cite{Booth:2022vwo} which is a generalization and systemization 
of that used in \cite{Booth:2021sow}. 

The formalism assumes a rotational symmetry of $h_{ij}$ and $K_{ij}$ generated by a coordinate vector field
$\partial_\phi$. It then identifies axisymmetric MOTS in $\Sigma_T$ by manipulating \eqref{eq:MOTS} into a 
pair of coupled differential equations on the (two-dimensional) orbit space $\tilde \Sigma := \Sigma_T / SO(2)$.
The equations describe an accelerated curve:
\begin{equation}
    \hat{T}^b \underline{\nabla}_b \hat{T}^a = \kappa_{\textrm{\tiny{MOTS}}} \hat{N}^a. \label{eqn:TDT}
\end{equation} 
where $\hat{T}$ and $\hat{N}$ are the unit tangent and normal vectors to the curve in $\tilde{\Sigma}$ and
$\kappa_{\textrm{\tiny{MOTS}}}$ is the magnitude of the acceleration. Solutions of these equations are dubbed \emph{MOTSodesics} and can be rotated by $\partial_\phi$ into 
full MOTS in $\Sigma_T$.   

The main work of the formalism is calculating $\kappa_{\textrm{\tiny{MOTS}}}$. For non-rotating spacetimes,
it is relatively straightforward. For rotating ones it can be quite involved. Luckily, we are dealing with a non-rotating spacetime but nevertheless, readers who are only interested in the final form of the equations may want to skip directly to \eqref{Peqn} and \eqref{Theqn}.

On the $\Sigma_T$ surfaces of constant $T$ the induced metric is
\begin{align}
    h_{ij} \dd x^i \dd x^j = N^2\dd X^2 + r^2 \dd \Omega^2 \label{eq:h}
\end{align}
and the extrinsic curvature is 
\begin{equation}
K_{ij} \dd x^i \dd x^j = N_T \dd X^2 + \frac{r r_T}{N} \dd \Omega^2 \; . 
\end{equation}
The is most easily calculated by recalling that 
$K_{ij} = \frac{1}{2} \mathcal{L}_{\hat{u}} h_{ij}$ and applying
\begin{equation}
\hat{u} = \frac{1}{N}  \frac{\partial}{\partial T}  \, ,   \label{u}
\end{equation}
where $N$ and $r$ were defined in \eqref{eqn:N} and \eqref{eqn:r} respectively. 
Clearly $\partial_\phi$ is a symmetry of both $h_{ij}$ and $K_{ij}$ and, with $r=r(T,X)$, both are explicitly 
time dependent.

% We can therefore employ the general framework~\cite{Booth:2022vwo} for constructing axisymmetric MOTS by describing them as curves in the orbit space $\tilde \Sigma := \Sigma / SO(2)$.  In concrete terms, both the intrinsic and extrinsic geometry of these hypersurfaces will depend on the time slice chosen, labelled by a parameter $T$, and the coordinates $(X,\theta)$. 

% The timelike unit normal one-form field 
% \begin{equation}
% u_\alpha = -N [\td T]_\alpha \quad \mbox{where} \quad N:=\left[\frac{32M^3 e^{-\frac{r}{2M}}}{r} \right]^{1/2}.
% \end{equation} 
% with orientation chosen so that 
% \begin{equation}
% u^\alpha = N^{-1} \left[ \frac{\partial}{\partial T} \right]^\alpha \, ,  \label{u}
% \end{equation}
% is forward-in-time pointing. 

Following the formalism of~\cite{Booth:2022vwo} we fix coordinates $x^a = (X,\theta)$ with the remaining coordinate $\phi$ representing the symmetry direction. 
Then the orbit space $(\tilde\Sigma, \uh_{ab},\underline{\nabla})$ has metric
\begin{equation}\label{eqn:uhab}
    \uh_{ab} \td x^a \td x^b = N^2 \td X^2 + r^2 \td \theta^2 \, , 
\end{equation} 
%$N = N(r)$, and $r$ is to be understood as a function of $X$ and the parameter $T$. 
and the Christoffel symbols associated to the metric connection $\underline{\nabla}$ are 
% \begin{equation}
% \Gamma^X_{~XX} = \frac{(N^2)' r_X}{2N^2}, \quad \Gamma^X_{~\theta\theta} = -\frac{r r_X}{N^2}, \quad \Gamma^{\theta}_{~X\theta} = \frac{r_X}{r},
% \end{equation}
\begin{equation}
\underline{\Gamma}^X_{~XX} = \frac{N_X}{N}\, , \; \;  \underline{\Gamma}^X_{~\theta\theta} = -\frac{r r_X}{N^2} \, 
\mbox{and} \; \;  \underline{\Gamma}^{\theta}_{~X\theta} = \frac{r_X}{r}, \label{eqn:Chr}
\end{equation}
where a subscript $X$ denotes partial differentiation with respect to $X$. 
 
% In concrete terms we are looking for a curve in the $(X,\theta)$ plane that is parameterized by arclength as:
% \begin{equation}
%     X = P(s) \quad \mbox{and} \quad \theta = \Theta(s) \; . 
% \end{equation}

We consider unit speed curves $x^a(s) = (P(s), \Theta(s))$ in $\tilde{\Sigma}$. These have tangent vector field
\begin{equation}
   \hat{T} = \dot P \pdv{X} + \dot \Theta \pdv{\theta},
\end{equation} 
where the overdot denotes differentiation with respect to the arc length parameter $s$. The unit speed condition imposes the constraint 
\begin{equation}
    N^2 \dot P^2 + r^2 \dot \Theta^2 =1.
\end{equation} 
The corresponding unit normal vector field along this curve is
% \begin{equation}
%     \hat{N} = \frac{1}{Nr}\left[ r^2 \dot \Theta \pdv{X} - N^2 \dot P \pdv{\theta} \right].
% \end{equation}
\begin{equation}
    \hat{N} = \left( \frac{r }{N}  \right)  \dot{\Theta} \pdv{X} - \left( \frac{N}{r} \right)  \dot{P} \pdv{\theta} \; \label{eq:hatN}.
\end{equation}

Then from section 2 of \cite{Booth:2022vwo} the acceleration 
%a two-dimensional MOTS in the hypersurface $(\Sigma, h_{ij}, K_{ij})$ is generated by a smooth curve $x^a(s)$ in $(\tilde\Sigma, \uh_{ab})$ 
%satisfying 
% the MOTSodesic equations are
% \begin{equation}
%     \hat{T}^b \underline{\nabla}_b \hat{T}^a = \kappa_{\textrm{\tiny{MOTS}}} \hat{N}^a.
% \end{equation} 
% Here 
\begin{equation}
\kappa_{\textrm{MOTS}}:= \mathcal{K} + \mathcal{K}_{\hat N}+ \mathcal{K}_{\hat{T}\hat{T}},
\end{equation} where
\begin{equation}
\begin{aligned}
    &\mathcal{K}:=h^{\phi\phi} K_{\phi\phi} \; , \\
    & \mathcal{K}_{\hat{N}} :=\hat{N}^a \underline{\nabla}_a (\ln \sqrt{h_{\phi \phi}}) \quad \mbox{and} \\
    & \mathcal{K}_{\hat T\hat T}:= K_{ij}\hat{T}^i \hat{T}^j \; . 
\end{aligned}\end{equation} 
% and $\underline{\mathcal{K}}_{ab}$ denotes the restriction of the extrinsic curvature $K_{ij}$ to vector fields tangent to $\tilde{\Sigma}$, i.e. $\pdv{x^a} = (\pdv{X}, \pdv{\theta})$. Then in this case 
Explicitly these are
\begin{equation}
    \begin{aligned}
        \mathcal{K} &= \frac{r_T}{N r} \\
        \mathcal{K}_{\hat N} &= \frac{\dot \Theta r_X}{N} - \frac{N \cot \Theta \dot P}{r} \quad \mbox{and} \\
       % \underline{\mathcal{K}}_{\hat{T} \hat{T} } &=\frac{r_T}{N} \left( \frac{(N^2)' \dot P^2}{2} + r \dot \Theta^2\right) \, , \\
        \underline{\mathcal{K}}_{\hat{T} \hat{T} } &= N_T \dot{P}^2 +  \frac{r r_T}{N} \dot{\Theta}^2 \, , 
    \end{aligned}
    \label{eqn:cK}
\end{equation} 
where a subscript $T$ denotes a partial derivative with respect to $T$.

% and %To compute the final term, note that the extrinsic curvature tensor can be written
% % \begin{equation}
% %     K_{ij}\td x^i \td x^j = \frac{r_T}{N}\left(\frac{(N^2)' \td X^2}{2} + r\td\Omega^2 \right),
% % \end{equation} so that we read off
% \begin{equation}
% \underline{\mathcal{K}}_{\hat{T} \hat{T} }  =\frac{r_T}{N} \left( \frac{(N^2)' \dot P^2}{2} + r \dot \Theta^2\right) . 
% \end{equation} 
Then by \eqref{eqn:TDT}, \eqref{eqn:Chr} and \eqref{eqn:cK} we obtain the MOTSodesic equations:
% \begin{align}
%     \ddot P & = -\frac{(N^2)' r_X \dot P^2}{2N^2} + \frac{r r_X \dot \Theta^2}{N^2} + \kappa_{\text{MOTS}}\frac{r \dot \Theta}{N}, \label{Peqn} \\
% \ddot \Theta & = -2\frac{r_X \dot P \dot \Theta}{r} - \kappa_{\text{MOTS}} \frac{ N \dot P}{r}, \label{Theqn}
% \end{align} where
\begin{align}
    \ddot P & = -\left(\frac{N_X }{N}\right) \dot P^2 + \left(\frac{r r_X}{N^2}\right)  \dot{\Theta}^2+ \left(\frac{r\kappa_{\text{\tiny{MOTS}}} }{N}\right) \dot \Theta \, , \label{Peqn} \\
\ddot \Theta & = -\left( \frac{2r_X }{r} \right) \dot P \dot \Theta - \left(\frac{ N \kappa_{\text{\tiny{MOTS}}}  }{r} \right)
\dot P, \label{Theqn}
\end{align}
where 
% \begin{equation}
%     \begin{aligned}
%     \kappa_{\text{\tiny{MOTS}}} &= \frac{r_T}{N}\left(\frac{1}{r} + \frac{(N^2)' \dot{P}^2}{2} + r \dot \Theta^2\right) + \frac{r_X \dot \Theta}{N} \\ & \quad - \frac{N \dot P \cot \theta}{r}.
%     \end{aligned}
% \end{equation} 
\begin{equation}
    \begin{aligned}
    \kappa_{\text{\tiny{MOTS}}} = & \frac{r_T}{rN} - \left(\frac{N \cot \Theta}{r} \right) \dot{P} + \left( \frac{r_X}{N} \right) \dot{\Theta} \\
    & + (N_T)  \dot{P}^2 + \left(\frac{r r_T}{N} \right) \dot{\Theta}^2.
    \end{aligned}
\end{equation}

As a simple check, consider a cross-section of the event horizon branch defined by $X = T= \mbox{constant}$. We then have $\dot P =0$ and the unit speed condition implies $\dot\Theta = \pm (2M)^{-1}$ so that $\ddot \Theta =0$. Choosing the upper sign for concreteness, note that \eqref{Theqn} is automatically satisfied. We then find \eqref{Peqn} reduces to the requirement 
\begin{equation}
    \frac{r_X + r_T}{M N^2} =0
\end{equation} which holds automatically since $r_X = 4M e^{-1} X = -r_T$ when evaluated on the surface $T = X$. This verifies that the cross-sections of the event horizon are indeed MOTS. 

% \ktb{Billy finds this extrinsic curavture instead:}
% \begin{equation}
%     K_{ij}\td x^i \td x^j = \frac{r_{\footnotesize{T}}}{r\sqrt{N}}\left(\frac{r~N' \td X^2}{2} + r^2\td\Omega^2 \right).
% \end{equation}
% \begin{multline}
% K_{ab}\dd x^a\dd x^b=&\sqrt{\frac{r^3}{32M^3}}r_{\scriptscriptstyle T} e^\frac{r}{4M}\left[ -\frac{16M^3}{r^3}e^\frac{-r}{2M}\left( 1+\frac{r}{2M} \right)\dd X^2+\dd\Omega^2 \right]
% \end{multline}

% \ktb{Should discuss the orientation of MOTS$<$-$>$MITS as one goes between $X>0$ and $X<0$ (there is an linear(?) dependence on $X$ in the MO(I)Tsodesic equations that makes it MITsodesics when the sign of $X$ flips.)}

\subsection{Visualizing a MOTS}

We visualize the MOTS by plotting their corresponding MOTSodesics in $\tilde{\Sigma}$. However, we first need to choose a way to represent $\tilde{\Sigma}$. We use two different methods, each one with advantages and disadvantages.

% With $- \infty < X < \infty$ along with an angle $\theta\in(0,\pi)$ there isn't a perfect way to to do this. Hence we use two different methods, each of which has strengths and weaknesses. 

% Given that 
% need to choose a visualization of $\tilde{\Sigma}$ given that $- \infty < X < \infty$ and $0 < \theta < \pi$. We do this in 
% two ways.

\begin{enumerate}[1)]
\item \emph{Polar-like coordinates:} The first is to map $\tilde{\Sigma}$ into the half-plane via $(x,y)=(e^X\cos\theta,~ e^X\sin\theta)$. In these coordinates, $X\to -\infty$ maps to the origin $(0,0)$  while 
$X\to\infty$ sends 
$x,y\to\infty$. The throat of the wormhole ($X = 0$) is the unit semi-circle in these coordinates. 

This method gives a simple, two-dimensional representation of the MOTS (see, for example, Figure \ref{fig:torus1}). However, the coordinate system distorts the geometry. While the two asymptotic ends of $\tilde{\Sigma}$ are geometrically equivalent, they appear very differently in this coordinate system. 

% However, this coordinate system does not treat the two asymptotic ends of $\tilde{\Sigma}$ 

% This representation has the advantage of being planar and relatively easy to plot (see, for example, Figure \ref{fig:torus1}). The disadvantage is that it hugely distorts the geometry: while two asymptotic ends of $\tilde{\Sigma}$ are geometrically equivalent, they are appear very differently in the diagram. 

% \ivan{Ivan: Not sure what you were doing in the second option. Is the $\rho$ supposed to be an $r$? I would have 
% expected this to be an embedding with $z=X$ and $r=r(X)$ but that doesn't appear to be what you did? Or is there a typo 
% of some kind? \ktb{Hi Ivan, I've added some more technical details to this section now. Hopefully it will be more clear and feel free to change things if you don't like the formatting (I have a feeling you won't like the $\xi$).}}

\item \emph{Embedding}: The second is to represent $\tilde{\Sigma}$ 
as an embedded surface in the half
of Euclidean $\mathbb{R}^3$ that is covered by 
cylindrical coordinates
$0 < \rho <\infty,-\infty< z < \infty$ and 
$0 < \vartheta < \pi$. Note that this is different from standard embedding diagrams. One typically embeds the disk at fixed $\theta=\pi/2$ on the full Euclidean $\mathbb{R}^3$. However, the interesting features of the MOTSs are encoded in the $\theta$ coordinate, so embedding the orbit space $\tilde{\Sigma}$ better showcases their properties.

The metric is 
\begin{equation}
\label{eqn:euclidean}
\td s^2=\td z^2+\td \rho^2+\rho^2\td\vartheta^2 \;  .
\end{equation}
A surface parameterized by
$z = z(X)$ and $\rho=\rho(X)$ then has metric
\begin{equation}\label{eqn:em2}
\td s^2=\left(\left(\dv{z}{X}\right)^2+\left(\dv{\rho}{X}\right)^2 \right) \td X^2+\rho^2\td\vartheta^2~.
\end{equation}
Matching with the induced metric
\eqref{eqn:uhab} 
on $\tilde{\Sigma}$ (holding $T$ constant such that $r(T,X)\to r(X)$) we obtain equations for the embedding:
\begin{align}
\vartheta & = \theta \; , \quad \rho = r(X) \quad \mbox{and} \\
N^2 & = \left(\left(\dv{z}{X}\right)^2+\left(\dv{\rho}{X}\right)^2 \right) \nonumber
 \; . 
\end{align}
On (numerically) solving the differential equation for $z(X)$ we have 
a parameterization of the embedding surface
in terms of $(X,\vartheta)$. Due to the symmetry about $X=0$, the solution for $z(X)$ is multiplied by a factor of $\mbox{sign}(X)$ to encode this symmetry about the $z=0$ plane. Figure \ref{fig:embedded1} shows the embedding diagrams of the same constant-$T$ slices that appeared in Figures~\ref{fig:kruskalDiagram} and~\ref{fig:penrosediagrams1} along with examples of MOTSs. 

The Einstein-Rosen bridge is clearly shown, along with the fact that it pinches off at $T = 1$.  The symmetry about $X=0$ has also been constructed to be shown about the $z=0$ plane. We emphasize that these diagrams do not reflect the axisymmetry (invariance under rotations in the periodically identified coordinate $\phi$). The MOTSs are ultimately the curves in $\tilde\Sigma$ (such as those depicted in Figures \ref{fig:torus1}, \ref{fig:ks0_time_evo}, \ref{fig:ks1_time_evo}, \ref{fig:loop-bifur}) rotated about the $\phi$ direction to result in surfaces such as Figure \ref{fig:DonutRotated}.

\begin{figure*}
\centering
\includegraphics[width=0.9\textwidth]{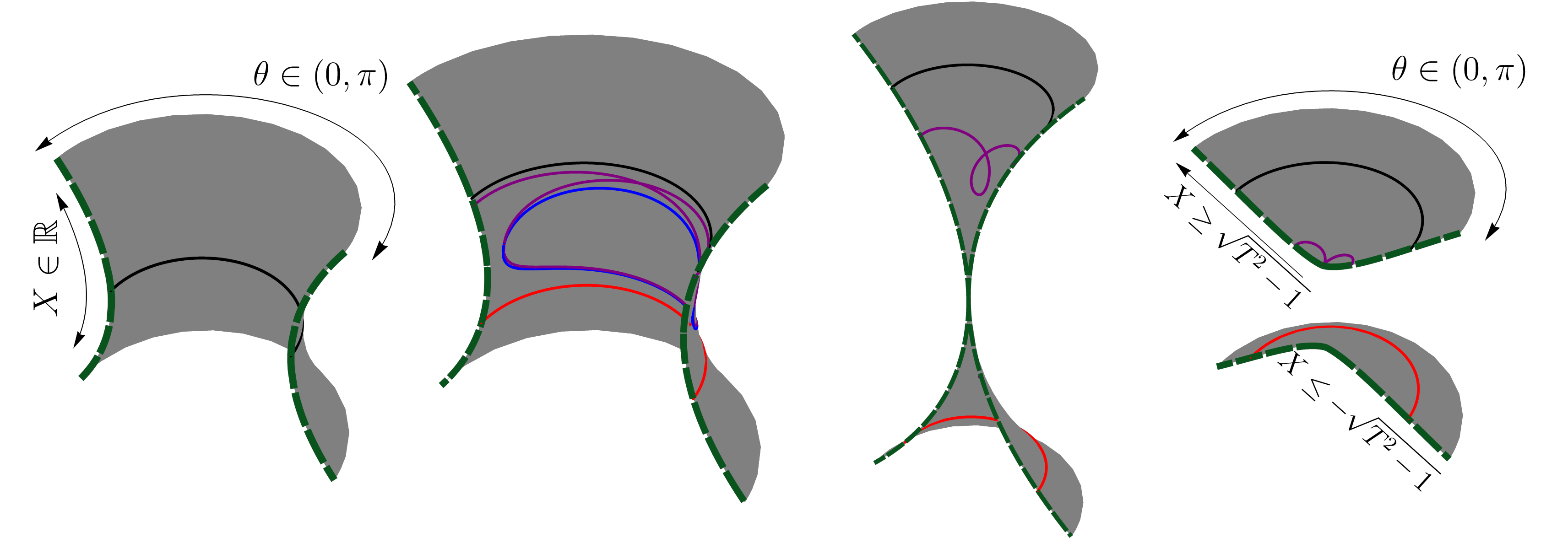}
    \caption{Embedding diagrams of $T=0.0,~0.5,~1.0,~1.5$ hypersurfaces, respectively from left to right. The solid black line is the $r=2M$ horizon at $X= T$ and the solid red is the $r=2M$ horizon at $X=-T$. The solid blue line is the toroidal MOTS at $T=0.5$ (in the family of MOTSs depicted in Figure \ref{fig:torus1}). The solid purple line is the once-intersected MOTSs in the respective $T$ slices (exemplary of the MOTSs shown in Figure \ref{fig:loop-bifur}). The dashed green lines are the poles at $\theta=0$ and $\theta=\pi$ in the usual Kruskal-Szekeres coordinate $\{T,X,\theta,\phi\}$. By mirroring these diagrams about $\theta=0,\pi$, one would recover the cartoon in Figure \ref{fig:EmbeddingPicture}.
    }
    \label{fig:embedded1}
\end{figure*}

% \begin{figure*}
% \centering
% \includegraphics[width=\textwidth]{figs/embed_final_v8_fullrotation_gray.pdf}
%     \caption{Embedding diagrams of constant $T=0.0,~0.5,~1.0,~1.5$, respectively from left to right. The solid black line is the $r=2M$ horizon at $X= T$ and the solid red is the $r=2M$ horizon at $X=-T$. The solid blue line is a once self-intersecting MOTS, the same shown in Figure \ref{fig:loop-bifur}.% \ktb{things to do: Text in figures could be larger, make the MITS red instead of black in accordance to other plots.}
%     }
%     \label{fig:embedded1}
% \end{figure*}

% \begin{figure*}
% \begin{center}
% \includegraphics[scale=.5]
% {figs/embed_final_v9_1_sideways_gray.pdf}~
% \includegraphics[scale=.5]
% {figs/embed_final_v9_2_sideways_gray.pdf}~
% \includegraphics[scale=.4]
% {figs/embed_final_v9_3_sideways_gray.pdf}~
% \includegraphics[scale=.5]
% {figs/embed_final_v9_4_sideways_gray.pdf}~
% \end{center}
%     \caption{Embedding diagrams of constant $T=0.0,~0.5,~1.0,~1.5$, respectively from left to right. The solid black line is the $r=2M$ horizon at $X= T$ and the solid red is the $r=2M$ horizon at $X=-T$. The solid blue line is a once self-intersecting MOTS, the same shown in Figure \ref{fig:loop-bifur}.% \ktb{things to do: Text in figures could be larger, make the MITS red instead of black in accordance to other plots.}
%     }
%     \label{fig:embedded1}
% \end{figure*}

\end{enumerate}

\subsection{Topology of a MOTS}

The MOTSs considered here are two-dimensional 
and orientable and so their 
topology is completely determined by the Euler characteristic. 
% To determine the topology of a given MOTS, we compute its Euler characteristic. 
This is found by integrating the Gauss curvature $K$ over the MOTS with induced area element $\dd a=\sqrt{\det(q_{AB})}~\dd s \wedge \dd\phi$,
\begin{equation}
    \chi=\frac{1}{2\pi}\int_S K \dd a~. \label{chi}
\end{equation}
The Gauss curvature $K=\frac{1}{2} R^{(2)}$, where $R^{(2)}$ is the scalar curvature associated to the induced metric on the MOTS
\begin{equation}
q_{AB} \dd x^A \dd x^B = \dd s^2 + q_{\phi\phi}(s)\dd \phi^2
\label{twometric}
\end{equation} where $s$ is the arc length parameter on the curve $(P(s), \Theta(s))$ which generates the surface. Explicitly, 
\begin{equation}
    q_{\phi\phi}(s) = r(T,P(s))^2 \sin^2 \Theta(s). \label{eq:qpp}
\end{equation}
In the present case, the induced metric is both diagonal and axisymmetric, so
\begin{equation}
R^{(2)}=-\frac{1}{(q_{\phi\phi})^{1/2}} \frac{\dd}{\dd s} \left(\frac{\dot q_{\phi\phi}}{(q_{\phi\phi})^{1/2}}\right) \label{eq:Ricci}
\end{equation}
The $(P(s),\Theta(s))$
from \eqref{Peqn} and \eqref{Theqn} determine $q_{AB}$
and $K$. Then in cases in which
the topology is unclear we can 
check the topology with \eqref{chi}.

% Our numerical construction of a MOTS produces $\left(P(s), \Theta(s)\right)$. The Ricci scalar can be computed from these interpolants and the integral performed over the surface, giving the Euler characteristic.

\subsection{The Stability Operator}
\label{sec:stab}

The stability operator encodes detailed information about a MOTS in its spectrum~\cite{Andersson:2005gq}. Generically, the spectrum of the stability operator is complex, but the principal (smallest) eigenvalue is guaranteed to be real. A MOTS is called strictly stable if its principal eigenvalue is positive, stable if it is non-negative, and unstable if it is negative. Strictly stable MOTSs enjoy a number of properties that make them well-suited to serve as quasi-local black hole boundaries. They are guaranteed to persist under time evolution and they serve as boundaries separating trapped and untrapped regions. 

While unstable MOTSs are unsuitable as quasi-local horizons they still play an important role in black hole dynamics. For example, the continuous sequence of MOTSs connecting the initial and final states in the head-on merger of two axisymmetric non-rotating black holes are largely made up from unstable MOTS~\cite{PhysRevLett.123.171102}. In particular unstable MOTSs are responsible for annihilating the apparent horizons of the original black holes in such a merger~\cite{Pook-Kolb:2021gsh}.  Here we will study the spectrum of the stability operator for the MOTSs found in the Kruskal slicing of the Schwarzschild black hole. 

Following the conventions of
\cite{Andersson:2005gq}
%the review~\cite{Chrusciel:2010fn}, 
the stability operator is
%\begin{equation}\label{staboperator}
%    L[\psi]:= -\Delta\psi + 2\mathcal{X}^i\partial_i \psi + \left[\frac{R^{(2)}}{2} -%\frac{1}{2} |\chi_+|^2 + \text{div} \mathcal{X} - |\mathcal{X}|^2\right]\psi
%\end{equation} 
\begin{align}
    L[\psi]:= 
       & -\Delta\psi + 2 \omega^A \partial_A \psi \label{staboperatorII} \\
    & + \left[\frac{R^{(2)}}{2} -\frac{1}{2} \|\sigma_+\|^2 + 
    \mathcal{\mathcal{D}}_A \omega^A  - \|\omega\|^2\right]\psi \nonumber \\
    = &  - \sD^A \sD_A \psi  \label{staboperatorIII} 
    + \frac{1}{2} \left[R^{(2)} -\|\sigma_+\|^2 \right]\psi  
\end{align} 
where $\| \sigma_+ \|^2 = \sigma^{AB}_+\sigma^+_{AB}$,  $\| \omega \|^2 = \omega^A \omega_A$ and $\sD_A = \mathcal{D}_A - \omega_A$. The shear and connection $\omega_A$ were defined in \ref{Sec:Geom} however for purposes of this 
section it is also useful to write them relative to initial data:
\begin{align}
\omega_A & = e_A^i K_{ij} \hat{N}^j \label{eqK} \\
\sigma^+_{AB} & = e^i_A e^j_B \left(K_{ij} + D_i \hat{N}_j \right) \; . 
\end{align}
Note that there is no need to subtract the trace to obtain a trace-free $\sigma^+_{AB}$. With
\begin{align}
\theta_+=  q^{AB} e^i_A e^j_B \left(K_{ij} + D_i \hat{N}_j \right)  = 0 \; , 
\label{thp}
\end{align}
it is automatically tracefree. 

In general the stability operator is not self-adjoint and has complex eigenvalues, however if $\omega$
is exact  then there is a significant 
simplification. This is most easily seen from the second form of the stability operator. If 
$\omega_A = \mathcal{D}_A \gamma$ for some $\gamma$, then for any scalar or tensor
$T$ a direct calculation shows that 
\begin{align}
\sD_A (e^\gamma T) = \left( \mathcal{D}_A  - \omega_A \right) (e^\gamma T) = e^\gamma \mathcal{D}_A T \;. 
\end{align}
It then follows that
\begin{align}
   \sD^A \sD_A  \psi = e^\gamma \mathcal{D}^A \mathcal{D}_A (e^{-\gamma}
   \psi) 
\end{align}
and so 
\begin{align} 
    L[\psi] = e^\gamma \bar{L} [e^{-\gamma} \psi ] \; . \label{rescale}
\end{align}
for 
\begin{align}
   \bar{L} :=  - \Delta + \frac{1}{2} \left[ R^{(2)} - 
    \| \sigma_+ \|^2 \right]  \; . \label{Lbar}
\end{align}
For further discussion of this transformation and its equivalence to rescaling the null
vectors, see \cite{Jaramillo_2015,BCM}. The transformation is significant as an operator of the form \eqref{Lbar}
is self-adjoint and so necessarily has a real eigenvalue spectrum with a smallest principal eigenvalue
$\lambda_0$. However, it is clear from \eqref{rescale} that if $L \psi_i = \lambda_i \psi_i$ then
$\bar{L} (e^{-\gamma} \psi_i) = \lambda_i (e^{-\gamma} \psi_i)$: they have the same eigenvalue spectrum
and the eigenfunctions of one are simple rescalings of those of the other. For our purposes we only 
care about the eigenvalue spectrum and so if $\omega_A$ is exact, we can forget 
about $\omega_A$ and instead study \eqref{Lbar}.

For the $\mathcal{S}$ generated by rotating
the MOTSodesic $(P(s), \Theta(s) )$ an application of \eqref{eq:hatN} and \eqref{eqK}  gives
\begin{equation}
    \omega = F(s) \td s \; \mbox{for } \;  F(s):= \dot P \dot\Theta r_T \left(\frac{1}{2} \frac{(N^2)'}{N^2} -1 \right) \; . 
\end{equation} 
This is exact and so henceforth we can forget about the complications of \eqref{staboperatorII} and instead calculate the eigenvalue spectrum of \eqref{Lbar}.

Calculating the rest of the terms in $\bar{L}$, the Laplacian for metric \eqref{twometric} is
\begin{equation}
    \Delta = \partial_s^2 + \frac{1}{2}\left(\frac{\dd}{\dd s}( \log q_{\phi\phi})\right)\frac{\partial}{\partial s} + \frac{1}{q_{\phi\phi}} \frac{\partial^2}{\partial \phi^2} \, ,
\end{equation} 
and the Ricci scalar was already calculated in \eqref{eq:Ricci}.
% Next, with $\hat{N}^i$ \eqref{eq:hatN} the unit vector in $\Sigma$ that points out of $\mathcal{S}$, we find that
% \begin{equation}
%     \omega = F(s) \td s , \quad F(s):= \dot P \dot\Theta r_T \left(\frac{1}{2} \frac{(N^2)'}{N^2} -1 \right).
% \end{equation} Then $|\mathcal{X}|^2 = F(s)^2$ and
% \begin{equation}
%     \text{div} \mathcal{X} = \frac{\dd}{\dd s} F + \frac{F}{2} \frac{\dd}{\dd s} ( \log q_{\phi\phi})
% \end{equation} 
The shear is a little more involved. Keeping in mind that $q_{AB}$ is diagonal, it follows from \eqref{thp} that
\begin{align}
\sigma^+_{ss} =-  q^{\phi \phi} \sigma^+_{\phi \phi} \; . 
\end{align}
Then, with $\sigma_{s \phi} = 0$ we have
\begin{equation}
    \|\sigma^+\|^2 = \frac{2 (\sigma^+_{\phi\phi})^2}{r^4 \sin^4\Theta} .
\end{equation}
where
\begin{align*} 
    % \sigma^+_{ss} & = N r (\dot P \ddot \Theta - \dot \Theta \ddot P) + 2 \dot P^2 \dot \Theta r_X N + \frac{r \dot \Theta^2}{N}(r_T + r_X r \dot\Theta) \\ & +\frac{\dot P^2 (N^2)'}{2N} (r_T - r r_X \dot\Theta)  \; \mbox{and} \\
     \sigma^+_{\phi\phi} & = \frac{\sin^2 \Theta}{N}\left(r r_T  + r^2 r_X \dot\Theta \right) -\cos \Theta \sin\Theta N r \dot P \; . 
\end{align*}

With these expressions in hand and given a numerical solution of \eqref{Peqn} and \eqref{Theqn}, we determine the spectrum of the stability operator numerically using pseudospectral techniques~\cite{Boyd, canuto2007spectral}. Since these methods have been described in detail elsewhere, e.g.~\cite{pook-kolb:2018igu, Hennigar:2021ogw}, we will be relatively brief with our overview. 

We are restricting here to MOTSs that share the axisymmetry of the space-time. Therefore, we expand the eigenfunctions of the stability operator as
\be 
\psi(s, \phi) = \sum_{m=-\infty}^{m = \infty} \psi_m(s) e^{i m \phi } \, ,
\ee
reducing the eigenvalue problem to a one-dimensional one. In the following we will restrict attention to the $m=0$ eigenfunctions since the principal eigenvalue, which determines the stability, must be invariant under the isometries of the MOTS.  

With the $\phi$-direction suppressed, the MOTS of interest reduces to an arclength parameterized MOTSodesic $\left(P(s), \Theta(s) \right)$ with $s \in [0, s_{\rm max}]$. We expand the eigenfunction of the stability operator in Chebychev polynomials,
\be 
\psi_m(s) = \sum_{n=0}^{N} a_n \cos \left(\frac{n \pi s}{s_{\rm max}}\right) \, ,
\ee
and divide the interval $[0, s_{\rm max}]$ into $N+1$ equally spaced points, 
\be
s_j=\frac{s_{\rm max}}{N+2}j \quad\mbox{for}\quad j\in\{1,2,\hdots,N,N+1\}.
\ee
Since the Chebychev polynomials are regular  at $s= 0$ and $s = s_{\rm max}$ it is not necessary to implement additional boundary conditions. 

Using the numerically determined MOTSodesic $\left(P(s), \Theta(s) \right)$ and the Chebyshev expansion of the eigenfunction, we construct a derivative matrix corresponding to the stability operator
\be 
\bar{L}[\psi] = \lambda \psi \to \bar{L}_{ij} a_j = \lambda \Phi_{ij} a_j \, ,
\ee
where $\bar{L}_{ij} = (\bar{L}_\Sigma \phi_j)(s_i)$ and $\Phi_{ij} = \phi_j(s_i)$. The spectrum of the stability operator is then determined by finding the eigenvalues of the matrix $\mathbf{M} = \mathbf{\Phi}^{-1} \mathbf{L}$, which we do using Mathematica. Convergence is tested by repeating the process for several distinct values of $N$.

\section{Toroidal MOTS\lowercase{s}} \label{sec:toroidal}

    \begin{figure*}
        \centering
        \includegraphics[width=.95\textwidth]{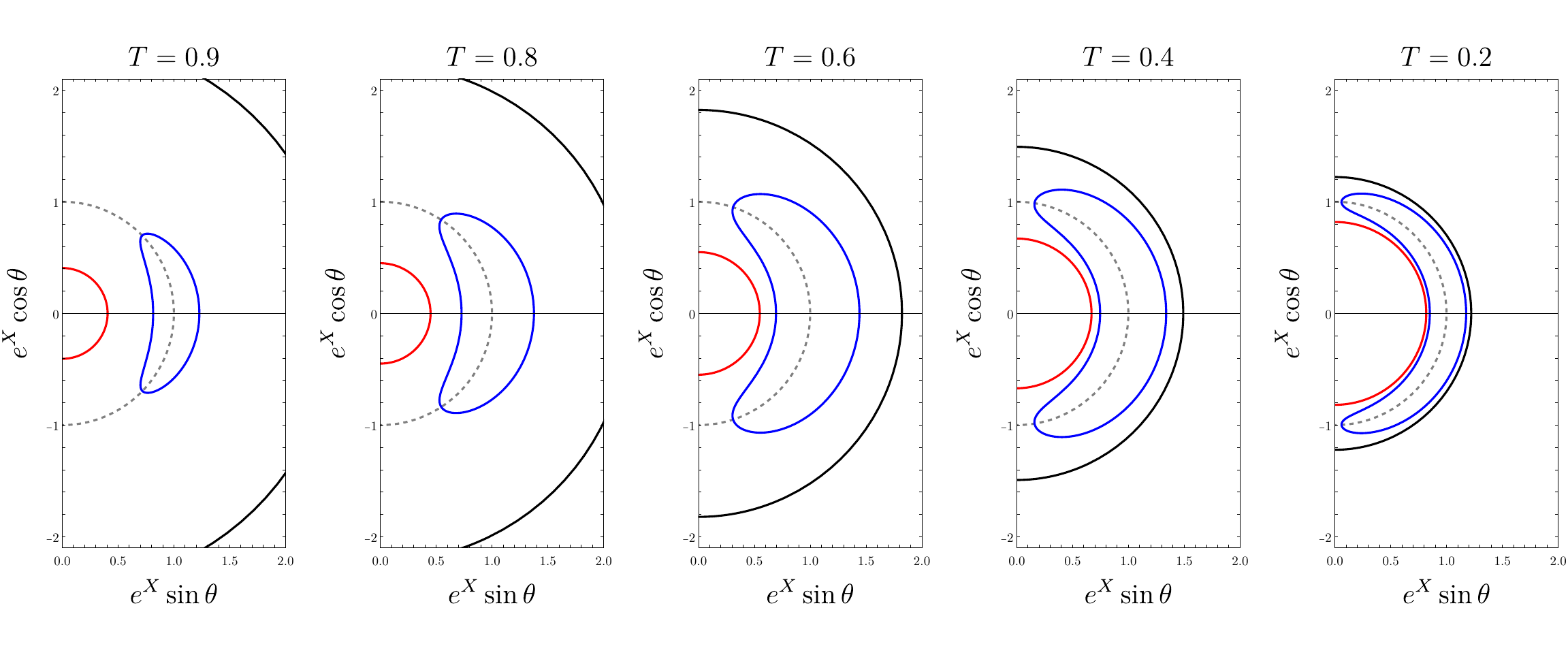}
        \caption{Toroidal MOTSs (in blue) found in the $T = 0.9$ to $T = 0.2$ hypersurfaces. The event horizon MOTS is shown in black ($X=T$) and MITS in red ($X=-T$). The place of spatial symmetry $X=0$ is shown as a dashed-gray line. Note the non-Euclidean axes.}
        \label{fig:torus1}
    \end{figure*}

The main result of our analysis is the observation of MOTSs of toroidal topology in the Kruskal-Szekeres time-slices. Toroidal MOTSs are located in the black hole interior and, in the examples we have found, the MOTS straddles the throat of the wormhole, meaning portions of the surface have $X > 0$ while other portions have $X < 0$. The extent of the toroidal MOTS in both asymptotic regions is equal, thus in these symmetrical cases we say the MOTS `straddles' the wormhole throat. In all cases, the numerically computed Euler characteristic is zero to within numerical precision.  None of the toroidal MOTSs we have located have self-intersections. 

We show several representative toroidal MOTSs in Figure~\ref{fig:torus1} for different values of Kruskal time. Numerically, we can resolve with confidence toroidal MOTSs from $0.145 < T < 1$. For $T < 0.145$ there are indications that these surfaces continue to exist, but their numerical identification is hampered due to a large number of nearby MOTSs. 

\begin{figure}
    \centering
    \includegraphics[width=.45\textwidth]{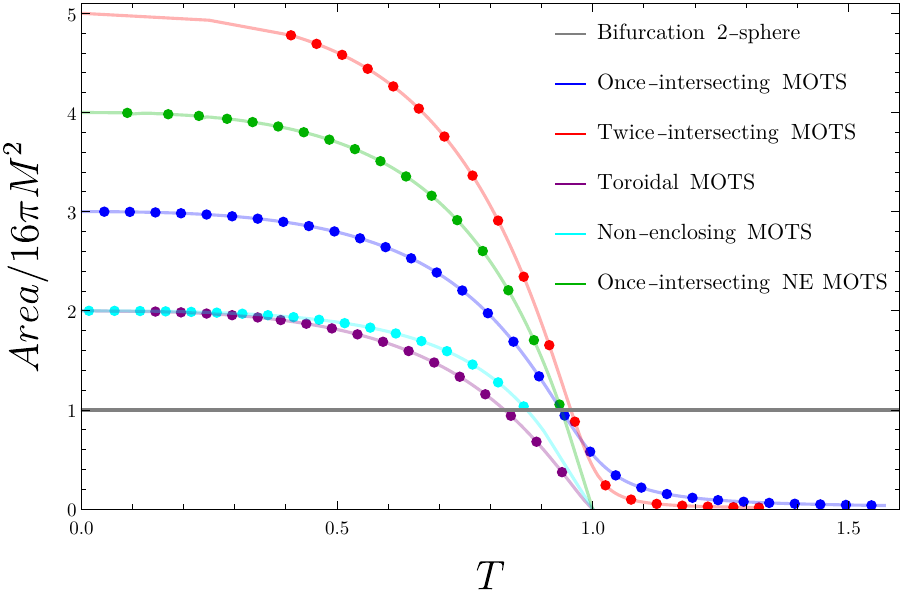}
    \caption{Areas of several families of MOTSs as a function of $T$. The dots correspond to numerical data, while the solid lines are interpolation and extrapolation of the numerical results. Note that not all MOTSs are shown to avoid overcrowding. The cyan and green class of MOTSs are introduced in Section \ref{sec:spherical}.}
    \label{fig:area_evo}
\end{figure}

As $T \to 1$, the time-slices approach the singularity.  We find that the toroidal MOTSs gradually shrink and at some finite time before $T = 1$ we no longer have the numerical accuracy to locate them. Moreover, we have found no examples of toroidal MOTSs for $T > 1$. This suggests that these surfaces are present only when Kruskal time-slices connect the left and right asymptotic regions via an Einstein-Rosen bridge. 

We can be more confident in this conclusion by studying the area of the MOTSs as a function of time $T$. The area of the MOTSs can be evaluated by integrating $\dd a=\sqrt{q_{\phi\phi}}~\dd s\wedge\dd\phi$ over the surface,
\begin{equation}
    \text{Area}=2\pi\int_s r(T,P(s))\sin(\Theta(s))~\dd s ~.
\end{equation} 
We show that the area of several MOTSs studied in this work in Figure~\ref{fig:area_evo}. The plot shows the area of the toroidal MOTSs is monotonically decreasing, approaching zero in the limit $T \to 1$.

The behaviour as $T \to 0$ is more subtle, since we can only resolve the torodial MOTS for $T > 0.145$. We find no clear indication that the toroidal MOTS annihilates with another surface. Instead, the resolution issue is related to the large number of MOTSs present in a smaller numerical domain. In the limit $T \to 0$, the MOTS and MITS corresponding to the intersection of the event horizon with the Kruskal time-slices become closer together. All interior MOTSs are sandwiched between the horizon MOTS/MITS, and distinguishing them from one another requires increasing precision as $T \to 0$. 

Tracking the area of this surface as $T \to 0$ we see that it approaches twice the area of the bifurcation two-sphere. This is consistent with the idea that as $T \to 0$, the toroidal MOTS becomes increasingly sandwiched between the horizon MOTS/MITS, effectively wrapping the bifurcation two-sphere twice as $T \to 0$. Confirming this qualitative picture will require improving the resolution of our numerical MOTS finder. 

Finally, we discuss the stability of the toroidal MOTSs. The toroidal MOTSs found in the numerical simulations performed in~\cite{Pook-Kolb:2021jpd} were all unstable with a negative principal eigenvalue. We find similar results here. Like that work, here we find that the shear $\sigma^+_{AB}$ is non-vanishing for the toroidal MOTSs.  This means that an outward space-like deformation of the MOTS does not lead to an untrapped surface, which is suggests these surfaces may be unstable. 

\begin{figure}
    \centering
    \includegraphics[width=.45\textwidth]{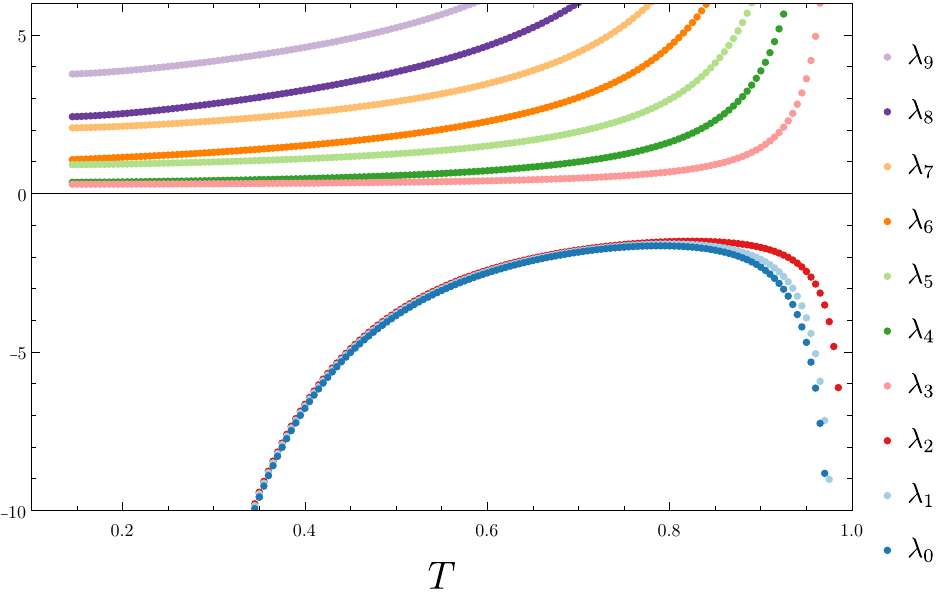}
    \caption{Lowest 10 eigenvalues of the stability operator on the toroidal MOTSs, as in Figure \ref{fig:torus1}, for different slices $T$. %\ktb{We resolve toroidal MOTSs from $T=0.145$ to $T=0.995$. As seen from the divergence of the first three eigenvalues, toroidal MOTSs will be harder to resolve for lower values of $T$. I could do a more `gradual' fade off for the left-tail of the plot. A less selective version is available uploaded here (`v1'). Some general comments: I would really like for $\lambda_3$ and $\lambda_4$ to approach 0.25, which is the principal eigenvalue of the bifurcation 2-sphere, which makes some sense to me since the sandwiching becomes tighter as $T\to0$. $\lambda_5$ similarly would approach 0.75 in this fantasy. All the negative eigenvalues ($\lambda_{(0-2)}$) would diverge as they have no place being an eigenvalue of the dreamy bifurcation two-sphere.}
    }
    \label{fig:toroidal_stability}
\end{figure} 

This expectation is confirmed via a direct evaluation of the spectrum of the stability operator. In Figure~\ref{fig:toroidal_stability} we plot the first few $m=0$ eigenvalues of the stability operator as a function of Kruskal time. The first three of these eigenvalues are strictly negative, confirming that the surfaces are unstable MOTSs. All other eigenvalues are positive. In the limit $T \to 1$, all eigenvalues appear to grow without bound. For smaller values of $T$, the first three eigenvalues take on large negative values. This is consistent with these eigenvalues diverging as $T \to 0$. The remaining eigenvalues remain finite as $T$ is decreased.

\section{Topologically Spherical MOTS\lowercase{s}} \label{sec:spherical}

In addition to the toroidal MOTSs, we find many other examples of MOTSs that all have spherical topology with numerically computed Euler characteristic $\chi = 2$. These MOTSs are reminiscent of the MOTS shown in Figure 12 of \cite{Pook-Kolb:2021jpd}. There are examples without and with self-intersections. 

This class of topologically spherical MOTSs do not enclose the $r=0$ curvature singularity. As such, we will synonymously refer to these MOTSs as the ``non-enclosing (NE)'' MOTSs. The toroidal MOTSs also do not enclose the singularity, but are uniquely identified with the `toroidal' name.

\begin{figure*}
    \centering
    \includegraphics[width=.95\textwidth]{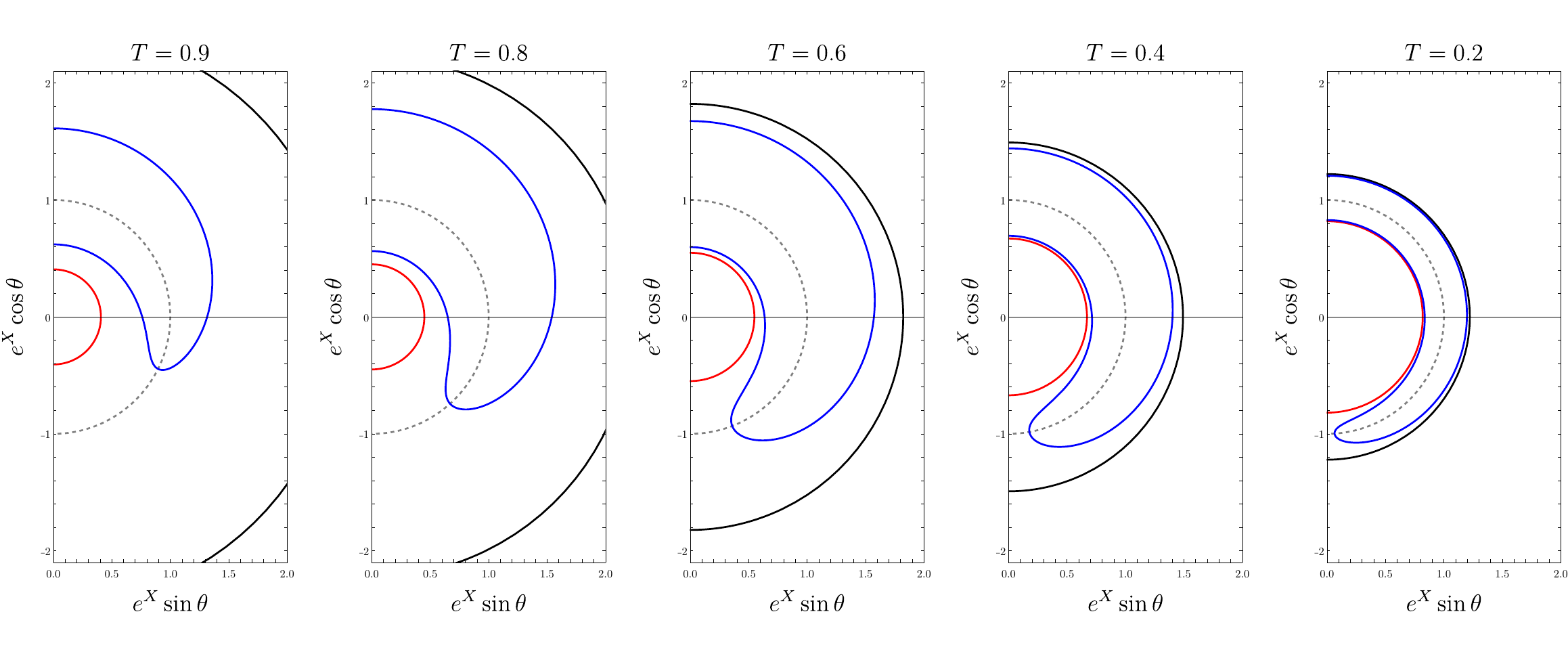}
    \caption{Non-intersecting non-enclosing MOTSs (in blue) found in the $T = 0.9$ to $T = 0.2$ hypersurfaces. The event horizon MOTS is shown in black ($X=T$) and MITS in red ($X=-T$). The place of spatial symmetry $X=0$ is shown as a dashed-gray line. Please mind the non-Euclidean axes.}
    \label{fig:ks0_time_evo}
\end{figure*}

Focusing first on examples without self-intersections, we present some examples of such surfaces in Figure~\ref{fig:ks0_time_evo}. These MOTSs are topological spheres with one pole occurring for $X > 0$ and the other occurring for $X < 0$. They are not geometric spheres and do not, for example, have constant Ricci curvature. %\ktb{These MOTSs are the only non-intersecting topologically spherical MOTSs found that are not the bifurcation 2-spheres.} 
%In the $(e^X \cos \theta, e^X \sin \theta)$ coordinates, these MOTSs look like a downward-curving `lip'.

MOTSs of this type can be located only for $T < 1$. As $T \to 1$, these MOTSs shrink in area and ultimately appear to vanish --- see the cyan curve in Figure~\ref{fig:area_evo} for the area as a function of time. On the other hand, as $T \to 0$, this MOTS becomes sandwiched between the horizon MOTS/MITS. While we ultimately cannot track this MOTS all the way to $T = 0$, its area evolution is suggestive of the fact that it ultimately wraps the horizon twice in this limit. 

\begin{figure}
    \centering
    \includegraphics[width=0.45\textwidth]{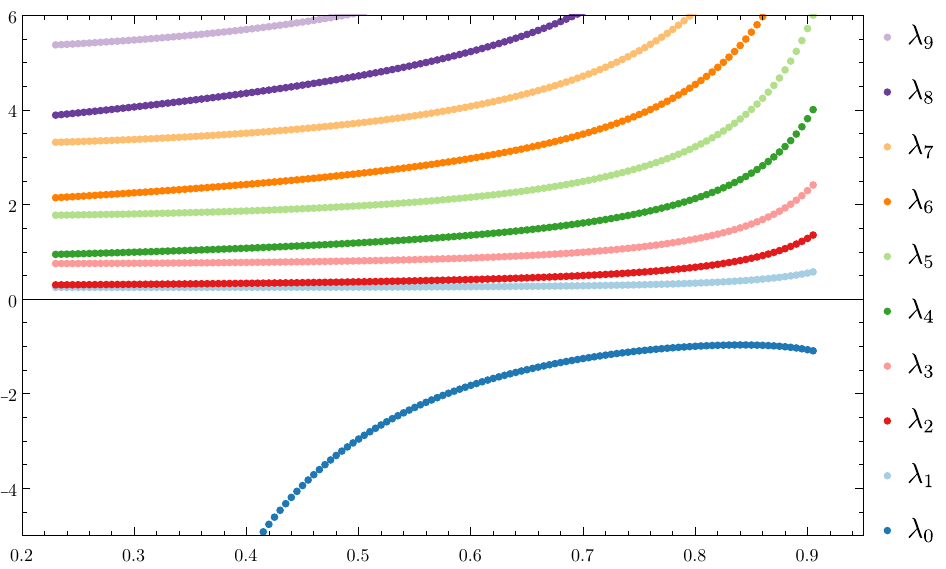}
    \caption{Lowest 10 eigenvalues of the stability operator on the topologically spherical but non-intersecting MOTSs (as in Figure \ref{fig:ks0_time_evo}) for different slices $T$.}
    \label{fig:lip0_stability}
\end{figure} 

These MOTSs are also unstable, as can be seen from their stability operator spectrum in Figure~\ref{fig:lip0_stability}. There is a single negative eigenvalue in the $m = 0$ sector of the spectrum. This eigenvalue tends toward $-\infty$ as $T \to 0$, while all other eigenvalue approach finite values. In the limit $T \to 1$ all eigenvalues appear to diverge.

\begin{figure*}
    \centering
    \includegraphics[width=.95\textwidth]{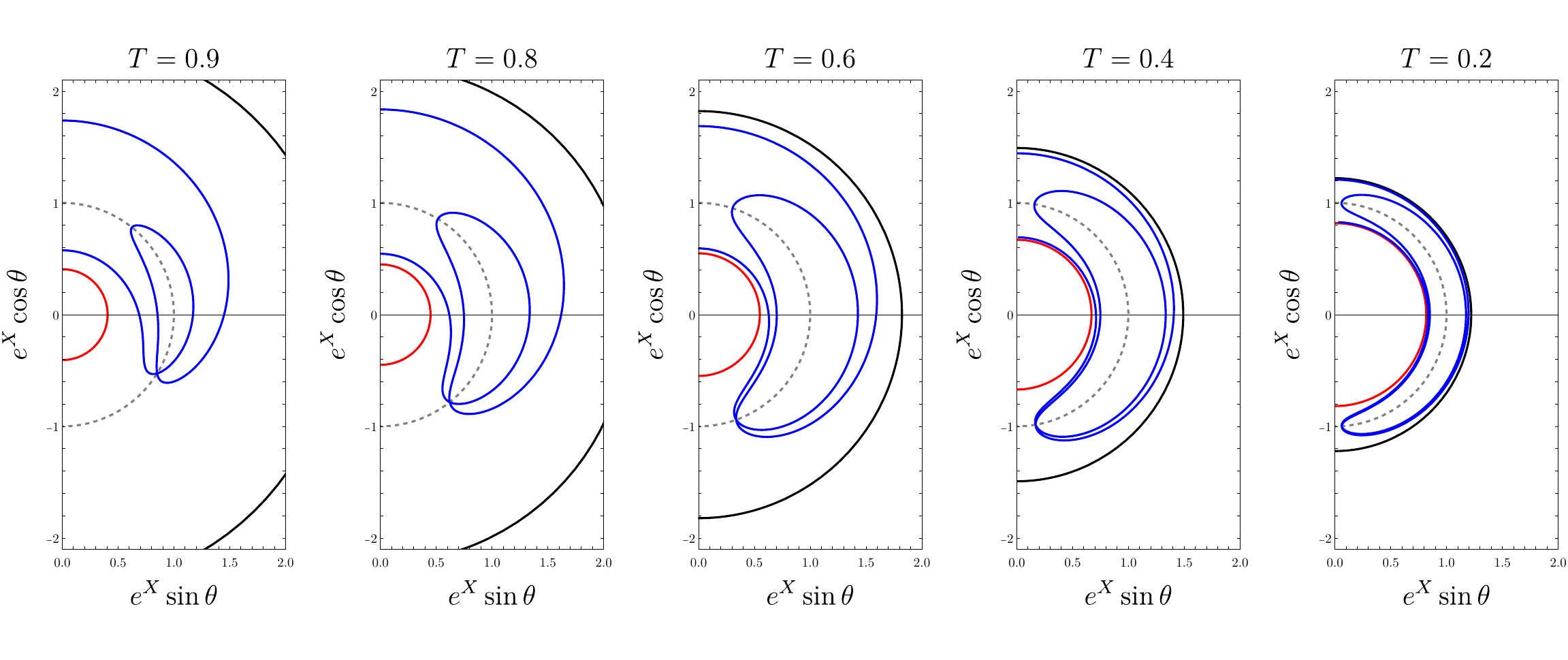}
    \caption{Once-intersecting non-enclosing MOTSs (in blue) found in the $T = 0.9$ to $T = 0.2$ hypersurfaces. The event horizon MOTS is shown in black ($X=T$) and MITS in red ($X=-T$). The place of spatial symmetry $X=0$ is shown as a dashed-gray line. Please mind the non-Euclidean axes.}
    \label{fig:ks1_time_evo}
\end{figure*}

\begin{figure*}
    \centering
    \includegraphics[width=.95\textwidth]{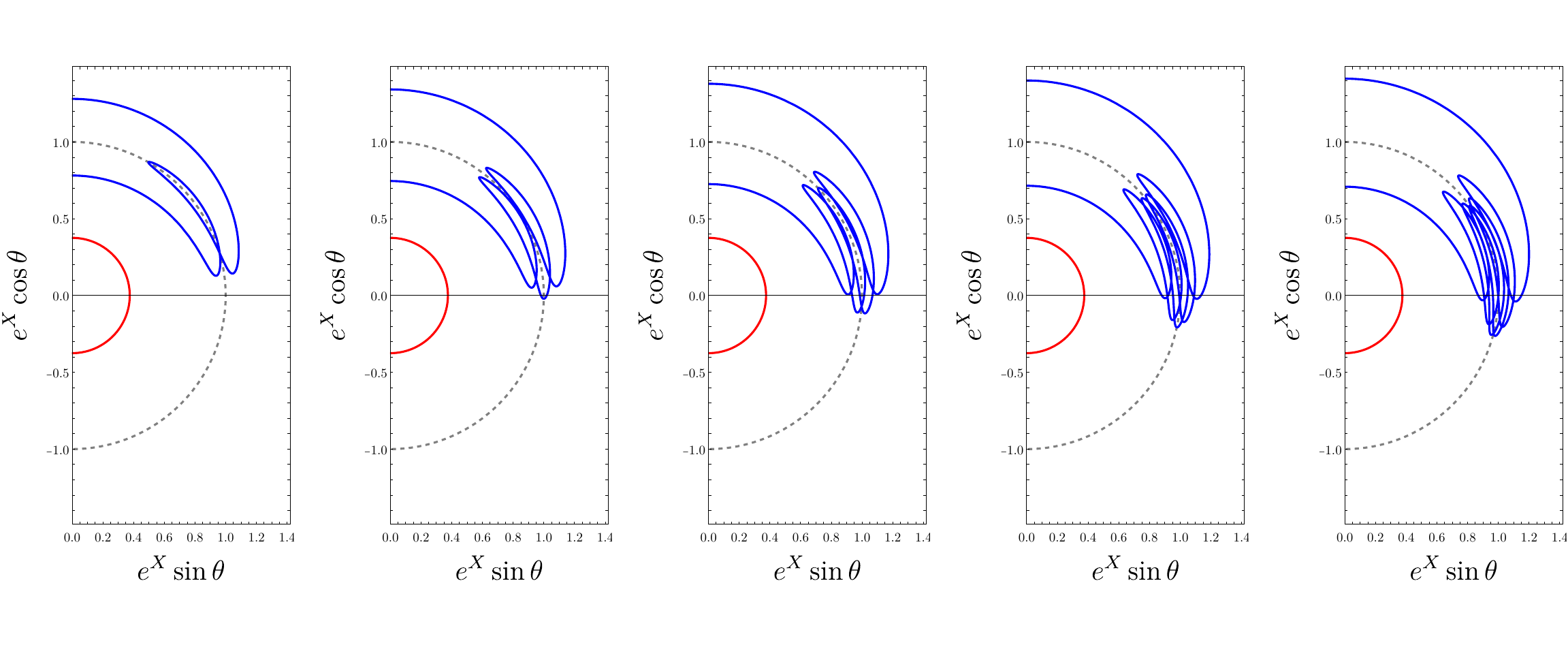}
    \caption{The once-, twice-, thrice-, four-times-, and five-times-intersecting non-enclosing MOTSs (in blue) found in the $T = 0.98$ hypersurface. The event horizon cross-section at $X=-T$ is shown in red --- the other sphere at $X=T$ is out of frame. The place of spatial symmetry $X=0$ is shown as a dashed-gray line. Please mind the non-Euclidean axes.}
    %The one, two, three, four, and five looping topologically spherical MOTSs at the $T = 0.98$ slice that do not enclose the $r = 0$ singularity. These also suggest that there exists a possible infinite amount of the N-looping MOTSs.}
    \label{fig:N_loopy_return_above}
\end{figure*}

Generalizations of these surfaces with self-intersections also exist, with examples shown in Figure~\ref{fig:ks1_time_evo}. These surfaces share the feature of having one pole at positive $X$ and the other at negative $X$. They span the throat of the wormhole, and can only be located for $0 < T < 1$. At Kruskal-Szekeres times near the collapse of the wormhole ($T=1-\varepsilon$), we are able to numerically resolve more self-intersecting non-enclosing MOTSs, shown in Figure~\ref{fig:N_loopy_return_above}. This suggests a large number of self-intersecting non-enclosing MOTSs that exists in the $T^2<1$ domain of hypersurfaces. We focus our attention only on the once-intersecting NE MOTSs in our stability operator analysis.

\begin{figure}
    \centering
    \includegraphics[width=0.45\textwidth]{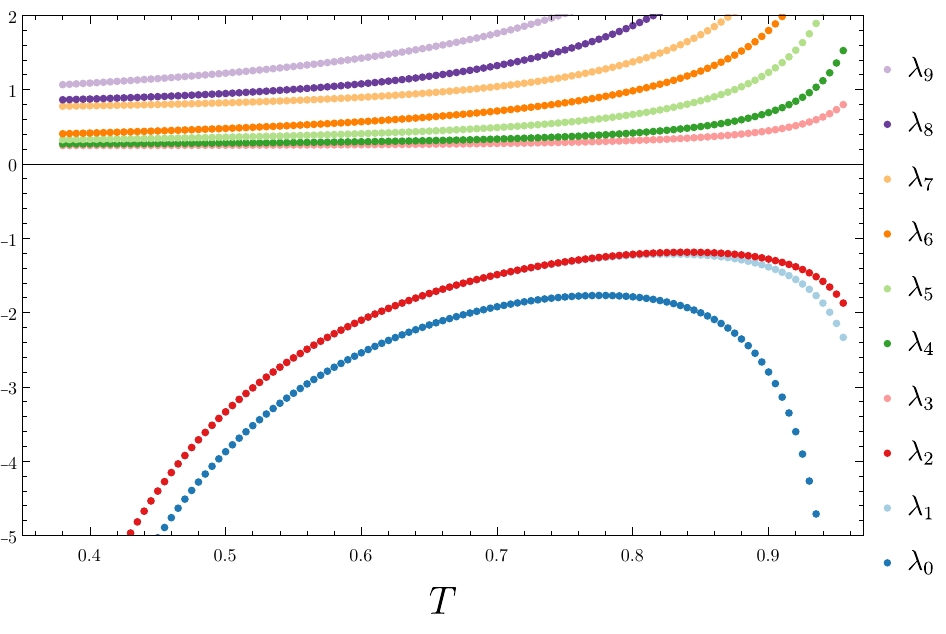}
    \caption{Lowest 10 eigenvalues of the stability operator on the once-intersecting MOTSs (such as in Figure \ref{fig:ks1_time_evo}) for different slices $T$.}
    \label{fig:lip1_stability}
\end{figure}

These MOTSs are also unstable, as can be seen in the stability operator eigenvalues plotted in Figure~\ref{fig:lip1_stability}. There are three negative eigenvalues, two more than the corresponding surfaces with no self-intersections. All the eigenvalues appear to diverge as $T \to 1$, while only the negative eigenvalues appear to diverge in the limit $T \to 0$. In this latter limit, it appears that the MOTS approaches the bifurcation two-sphere, wrapping around it four times --- see the green curve in Figure~\ref{fig:area_evo}. 

\begin{figure*}
    \centering
    \includegraphics[width=\textwidth]{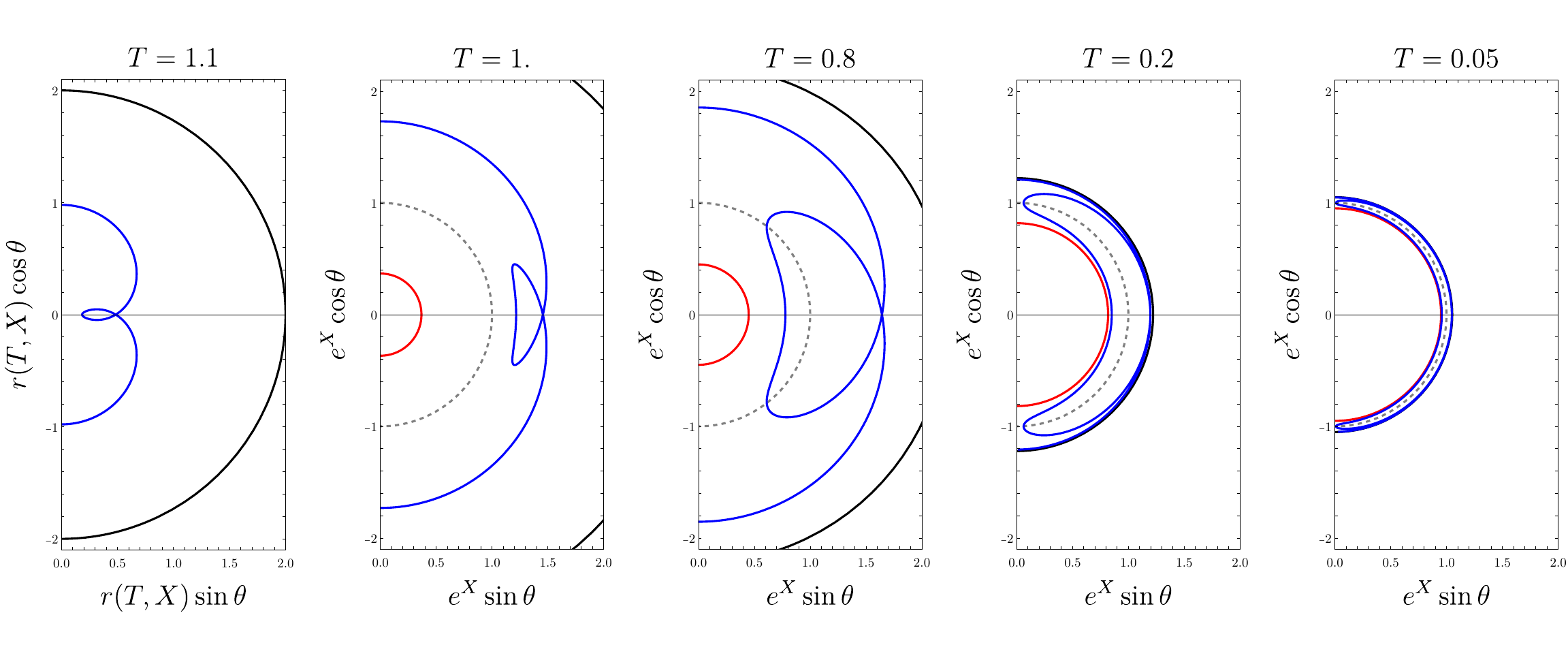}
    \caption{Once-intersecting MOTSs (in blue) found in the $T = 1.1$ to $T = 0.2$ hypersurfaces. The event horizon MOTS is shown in black ($X=T$) and MITS in red ($X=-T$). The place of spatial symmetry $X=0$ is shown as a dashed-gray line. Please mind the non-Euclidean axes other than the left-most panel. The left-most panel uses a Euclidean axes format to make clear the connection with~\cite{Booth:2020qhb}.}
    \label{fig:loop-bifur}
\end{figure*}

Finally, there are examples of self-intersecting MOTSs that are similar to those observed in Painleve-Gullstrand time-slices of the Schwarzschild space-time~\cite{Booth:2020qhb}. We show a time-progression for one of these MOTSs in Figure~\ref{fig:loop-bifur}. These MOTSs are topologically spherical, and have both their poles located at positive $X$ values. These are also the only MOTSs we have discussed so far that can be tracked beyond $T = 1$ --- see the blue curve in Figure~\ref{fig:area_evo}. We find examples of MOTSs of this type with multiple self-intersections, Figure~\ref{fig:loop-bifur} plots the time-progression for the once-intersecting surface, while we include the area evolution of both the once- and twice-intersecting surfaces of this kind in Figure~\ref{fig:area_evo}.

For $T < 1$, as is clear from Figure~\ref{fig:loop-bifur}, a portion of the MOTS extends across the wormhole at early times. As time increases toward $T = 1$, this MOTS pulls back until it is entirely contained in the $X > 0$ region. The MOTS can be continued to be tracked for larger values of $T > 1$, for which the time-slices terminate at the singularity. As $T$ continues to become larger, the MOTS becomes increasingly distorted and eventually can no longer be numerically resolved. See Figure~\ref{fig:embedded1}, which illustrates some of these features on an embedding diagram. 

On the other hand, as $T \to 0$ these MOTSs become sandwiched between the two components of the event horizon. This is shown in Figure~\ref{fig:loop-bifur} and is similar to what happens with all the MOTSs described earlier. Just like in that case, it is not possible to numerically resolve what happens in the strict $T \to 0$ limit, but tracking the area of the MOTS as a function of time is suggestive of the fact that it limits to a MOTS that wraps the bifurcation sphere some number of times. The number of wrappings appears to be equal to $2L+1$, where $L$ is the number of loops formed by self-intersection. It would be interesting to see if these new types of MOTSs arise generically in maximally extended black hole spacetimes.

\begin{figure}
    \centering
    \includegraphics[width=0.45\textwidth]{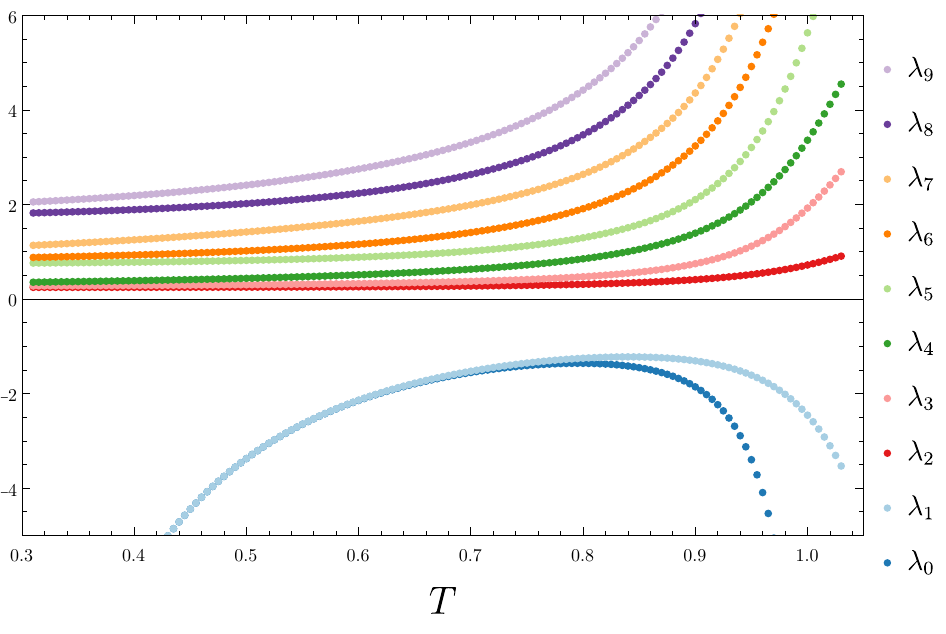}
    \caption{Lowest 10 eigenvalues of the stability operator on the once-intersecting MOTSs (as in Figure \ref{fig:loop-bifur}) for different slices $T$.}
    \label{fig:1_intersect_stability}
\end{figure} 

Just as in the Painlev\'e--Gullstrand slicing, these looping MOTSs are unstable. The $m=0$ eigenvalues of the stability operator are shown in Figure~\ref{fig:1_intersect_stability} for the once-intersecting MOTS. There are two negative eigenvalues, and both of these appear to be divergent in the $T \to 0$ limit. The positive eigenvalues appear to have finite limits as $T \to 0$. As $T$ increases, a number of eigenvalues begin to grow large, suggesting they ultimately diverge. However, the time at which this divergence occurs is no longer equal to $T =1 $, as these MOTSs can be be tracked beyond that point. There may be a connection between the loss of the ability to numerically resolve this once-looping MOTS and the divergence of its eigenvalue spectrum. As $T$ increases, the time-slices tend increasingly toward being null. A similar feature was observed for these MOTSs in the generalized Painlev\'e--Gullstrand slicing of the Schwarzschild space-time --- c.f. Figure 5 of~\cite{Hennigar:2021ogw}.

\section{Conclusions}

% \rah{Things to comment on:
% \begin{itemize}
%     \item Comment on the fact that the positive eigenvalues seem to approach those of the horizon as $T \to 0$, but with multiplicities in proportion to how many times that surface ``wraps'' the horizon.
%     \item \st{Cut out results for the MOTSodesic deviation approach. (Toroidal case ambiguous).}
% \end{itemize}}

% \ktb{
% \begin{itemize}
% \item Should scour through the figure captions, not necessarily in a publishable state on some of them (some where place holders when I had put them in)
% \item Introduced ``Non-Singularity-Enclosing (NE) MOTS''
% \item Added explanation of `straddle' in beginning of III.
% \end{itemize}
% }\\

We have investigated the existence of axisymmetric MOTSs with toroidal topology lying in constant $T$ spacelike hypersurfaces in the maximally extended Kruskal--Szekeres spacetime. This work is motivated by more sophisticated numerical simulations in \cite{Pook-Kolb:2021jpd} that reveal that toroidal MOTSs arise naturally in dynamical black hole mergers.  Their appearance in these examples suggests that toroidal MOTSs are not intrinsically associated with merger physics, but instead can be attributed to the departure from time-symmetry.

Along the way, our analysis demonstrated the existence of new spherical MOTSs in the Schwarzschild interior. These novel MOTSs are symmetric about the wormhole throat (i.e. about the timelike line $X=0$ in the standard Kruskal-Szekeres coordinates). In the region $T \in (-1,0)\cup(0,1)$, these MOTSs `straddle' the two asymptotic regions connected by the Einstein--Rosen bridge. For the hypersurfaces with $|T| > 1$, however, the Einstein-Rosen bridge collapses and the surface must intersect the singularity (as depicted in Figures \ref{fig:kruskalDiagram}, \ref{fig:penrosediagrams1}, and \ref{fig:embedded1}). 

The construction of these axisymmetric MOTSs is an application of the MOTSodesic method introduced in \cite{Booth:2020qhb, Hennigar:2021ogw, Booth:2022vwo}. These new types of MOTSs are unstable, as we demonstrated by a numerical analysis of the stability operator and its spectrum. In particular, the toroidal MOTSs appear to %typically
have three negative $m=0$ eigenvalues of the stability operator, whereas the topologically spherical MOTSs have a number of negative eigenvalues related to the number of self-intersections in a manner consistent with previous studies. These negative eigenvalues indicate that these MOTSs are not boundaries separating trapped and untrapped regions \cite{Andersson:2007fh}. 
%\rah{\st{Maybe re-iterate here one of the motivations: This shows that the these toroidal MOTSs is not associated with merger physics, but instead can be attributed as a departure from time-symmetry}}

% These toroidal and `lip' MOTSs that necessarily `straddle' the Einstein--Rosen bridge all vanish for $|T|\geq 1$, where the spacelike hypersurface intersects the spacelike $r=0$ singularity and separates for individual asymptotic regions $X^2\geq T^2-1$. The previously studied self-intersecting MOTSs (as depicted in FIGURE REF) progressively retreats from the wormhole throat as this separation occurs at $T^2=1$ and these MOTSs recover a similar form in the presence of the singularity. 

Finding MOTSs becomes difficult for $T\to 0$ and $T \gg 1$ (besides, of course, the MOTS corresponding to the event horizon). It can be shown that the constant $T$ spacelike hypersurfaces asymptote to null surfaces as $T$ increases. This is reflected by the divergence of eigenvalues of the stability operator. Near $T\to 0$, the density of MOTSs is substantial, as all found MOTSs are sandwiched between the two apparent horizons, and the problem becomes distinguishing them.

It is unclear whether the loss of MOTSs as $T\gg 1$ is due to them annihilating with other MOTSs at critical values of $T$. For example, `annihilation' or `bifurcation'  have been observed in previous studies \cite{Pook-Kolb:2021jpd,Hennigar:2021ogw} and occur simultaneously with a vanishing eigenvalue of the stability operator. As there is no evidence of any found eigenvalues going to zero together with eigenvalues diverging near the loss of the MOTS, we expect that these MOTSs are lost due to lack of numerical precision. Improving this issue is not trivial: simply increasing numerical precision will not definitively extend the range of our MOTS finding techniques.
There are a large number of axisymmetric MOTSs located in the Schwarzschild interior. As $T \to 0$, these are `sandwiched' into the increasingly small domain between the two apparent horizons and it becomes very difficult to distinguish individual MOTSs in this limit. Meanwhile, as $T\gg 1$, it has been previously observed how MOTSs behave for slices that transition from spacelike to null -- they steadily warp into a sharp surface that nears the $r=0$ singularity in the near-null coordinates. This behaviour is difficult to deal with numerically. 
% (3) Since the MOTSodesic equations are singular on the $z$-axis, a near-axis approximation was used (see CITE)
%It may be the case that closed-form solutions of the numerical MOTSs presented in this work can decisively shed light on the MOTSs behaviour at these extremities. \ivan{(Don't quite understand this sentence - what is the ``addresseed'' doing?).\ktb{ Changed it, maybe it is better now.}}\ivan{Better, but now I wonder about the phrase ``closed form" - often that is used as a synonym for ``exact solution". \ktb{Yes, Ivan, I think in this case it is referring to an exact (analytical) solution for the MOTSodesic equation.}}

The eigenvalue spectrum of the stability operator on the non-apparent horizon MOTSs exhibits a curious behaviour as $T \to 0$. We observe that the non-negative eigenvalues of the stability operator of all non-apparent-horizon MOTSs tends towards the corresponding eigenvalues of the event horizon cross-section, but with multiplicities. The multiplicity of each eigenvalue is equal to ratio of the area of the MOTSs to the area of the event horizon cross-section as $T \to 0$, which is an integer (Figure \ref{fig:area_evo}). As an example, the eigenvalue spectrum of the stability operator (Figure \ref{fig:lip1_stability}) for the once-intersecting NE MOTSs (Figure \ref{fig:ks1_time_evo}) showcases this. From the plots of its area as $T$ varies (Figure \ref{fig:area_evo}), the area of the once-intersecting NE MOTS approaches 4 times that of the event horizon cross-section. Towards the right-most panel of Figure \ref{fig:ks1_time_evo}, this MOTS `wraps' the event horizon cross-sections four times, hence the quadruple area factor. As this MOTS gets sandwiched between the two horizons, its unit normal vector $\hat{N}^a$ begins to coincide with that of the horizons with degeneracy due to the multiple wrapping. This picture is consistent with the multiplicities of the positive $m=0$ eigenvalues and the negative eigenvalues may be thought of as a result of the normal vector field pointing tangentially to the horizons. Investigating this feature may be a topic for future work.

It would be interesting to see whether similarly extended spacetimes would exhibit similar behaviour, for example a Kruskal-Szekeres extension of the Reissner--Norstr\"om spacetime. The MOTSs within the charged black hole have been studied in an earlier work \cite{Hennigar:2021ogw} using a further generalized Painlev\'e--Gullstrand slicing and exhibit the aforementioned annihilation/bifurcation events in the charge parameter space $Q$. Other coordinate systems that have a non-vanishing extrinsic curvature and that span both asymptotic regions, such as hyperboloidal slicings, could be studied further to see whether they harbour similar MOTSs to the ones found here.

\vfill

\section*{Acknowledgments} 
The author order of this paper is students (the first three authors) by degree of contribution, 
followed by more senior authors in alphabetical order by last name. \\
\indent We thank Graham Cox and Daniel Pook-Kolb for helpful comments. The work of RAH received the support of a fellowship from ``la Caixa” Foundation (ID 100010434) and from the European Union’s Horizon 2020 research and innovation programme under the Marie Skłodowska-Curie grant agreement No 847648 under fellowship code LCF/BQ/PI21/11830027. IB and LN were supported by Natural Science and Engineering Research Council of Canada (NSERC) Discovery Grant 2018-04873. HK was supported by NSERC Discovery Grant 2018-04887. SM and KTBS were supported by both of these NSERC Discovery Grants. KTBS acknowledges support from NSERC via a Postgraduate Scholarship -- Doctoral award. Memorial University (St.~John's campus) is situated on traditional territories of diverse Indigenous groups, and we acknowledge with respect the diverse histories and cultures of the Beothuk, Mi'kmaq, Innu, and Inuit of Newfoundland and Labrador. McMaster University is located on the traditional territories of the Mississauga and Haudenosaunee nations, and within the lands protected by the ``Dish with One Spoon'' wampum agreement.
 
\vfill

    % \begin{figure}
    %     \centering
    %     \includegraphics[width=0.25\textwidth]{figs/T=0.5MOTSodesic.pdf}
    %     \caption{MOTSodesic deviation of one-looping MOTS at $T=0.5$. Here, the solid blue line remains as the MOTS, where the green and red lines are MOTOS shot from positions very near the MOTS. This figure shows two intersections.}
    %     \label{fig:p5MOTSodesic}
    % \end{figure}

    % \begin{figure}
    %     \centering
    %     \includegraphics[width=0.3\textwidth]{figs/HighTMOTSodesic.pdf}
    %     \caption{MOTSodesic deviation of two-looping MOTS at $T=1.09$. Here, the solid blue line remains as the MOTS, where the green and red lines are MOTOS shot from positions very near the MOTS. This figure shows four intersections, which is expected for this two-looping MOTS.}
    %     \label{fig:highMOTSodesic}
    % \end{figure}

    % \begin{figure}
    %     \centering
    %     \includegraphics[width=0.3\textwidth]
    %     {figs/MOTSodesic_Tor.pdf}
    %     \caption{MOTSodesic deviation of torodial MOTS, where solid blue line is the toroidal MOTS and the purple and green lines are nearby MOTOS.}
    %     \label{fig:MOTSodesic_Tor}
    % \end{figure}

\pagebreak

\bibliography{blmotos}
    
%\printbibliography
		%	\input{char_ev_ArXiv_fix.bbl}
\end{document}